\documentclass[aps,prb,twocolumn,superscriptaddress,amsmath,amssymb,showpacs]{revtex4-1}
\usepackage{amssymb}
\usepackage{graphicx}
\usepackage{bm}
\usepackage{epsfig}
\usepackage{color}
\usepackage[ansinew]{inputenc}
\usepackage{amsmath}
\usepackage{ dsfont }
\usepackage{float}
\usepackage[table]{xcolor}
\usepackage{xcolor,colortbl}
\usepackage{tabularx}

\usepackage{tikz}
\usetikzlibrary{matrix}

\usepackage{youngtab}
\usepackage{young}

\newcommand{\dg}{^{\dagger}}
\definecolor{Gray}{gray}{0.9}
\definecolor{LightCyan}{rgb}{0.88,1,1}

\def\bk{{\bold{k}}}
\def\fb{{\bar{f}}}

\def\qb{{\bar{q}}}

\begin{document}

\title{Supersymmetric approach to heavy fermion systems}

\author{Aline Ramires}
\altaffiliation[Current address:]{ Institute for Theoretical Studies, ETH Zurich, Clausiusstrasse 47, Building CLV, 8092 Zurich, Switzerland.}
\affiliation{Department of Physics and Astronomy, Rutgers University,
Piscataway, New Jersey, 08854, USA}
\author{Piers Coleman}
\affiliation{Department of Physics and Astronomy, Rutgers University, Piscataway, New Jersey, 08854, USA}

\date{\today}

\begin{abstract}
We propose a generalization of the supersymmetric representation of
spins with symplectic symmetry, generalizing the rotation group of the
spin from $SU (2)$ to $SP(N)$. As a test application of this new representation, we
consider two toy models involving a competition of the Kondo effect
and antiferromagnetism: a two-impurity model and a frustrated
three-impurity model. Exploring an ensemble of L-shaped representations
with a fixed number of boxes in their respective Young tableaux, we
allow the system to choose which representation is energetically more
favorable in each region in parameter space. We discuss how the
features of these preliminary applications can generalize to Kondo
lattice models.
\end{abstract}

\maketitle

%%%%%%%%%%%%%%%%%%%%%%%%%%%%%%%%%%%%%%%%%%%%%%

\section{Introduction and motivation}

%Heavy Fermion Systems
Heavy fermion materials involve a lattice of localized magnetic
moments derived from f-electrons,  embedded in a conduction
sea formed principally from
delocalized d-electrons \cite{Hew, Col0}. The physics of these
materials can be understood as a consequence of the interplay between two
competing physical processes: 
\begin{itemize}
\item [-] the Kondo effect, which tends to screen
the local moments to produce a band of heavy electrons, and
\item [-] antiferromagnetism, which locks the local moments together
via the RKKY interaction, into a
state with long range magnetic order.  
\end{itemize}
The characteristic scales for
these two processes are 
\begin{eqnarray}
T_K \sim D e^{-1/\rho_0 J_K}, \hspace{1cm} T_{RKKY}\sim \rho_0 J^2_K,
\end{eqnarray}
where $J_{K}$ is the strength of the onsite Kondo interaction between
the localized f-electrons and conduction electrons, 
$\rho_0 \sim 1/D$ is the density of states of the
conduction electrons at the Fermi energy and $D$ the bandwidth. 
The Kondo temperature, $T_K$, sets the energy scale for the onset of
the Kondo effect and consequently the formation of a coherent heavy
Fermi liquid (HFL) while $T_{RKKY}$ defines the energy scale for
the onset of magnetic order.  

The family of heavy fermion materials provides an important setting
for the study of quantum criticality \cite{Si1, Geg} which develops
when a continuous second-order phase transition is suppressed to
absolute zero temperature.  The small characteristic energy scales of these
compounds makes them highly tunable, allowing the ready exploration of
the phase diagram as a function of pressure, magnetic field or
doping. Superconductivity is often found in the vicinity of magnetic
quantum critical points (QCP). At temperatures above the quantum
critical point non Fermi liquid (NFL) behavior is observed, generally
characterized by sub-quadratic temperature dependence of the
resistivity $\rho $ and a logarithmic temperature dependence of the
specific heat coefficient $\gamma= \frac{c_{V}}{T}$,
\begin{align}
\begin{split} \rho &\propto T^\alpha,\qquad (\alpha <2)\\
\gamma &\propto \frac{1}{T_{0}}\log \left(\frac{T_{0}}{T} \right),
\end{split}
\end{align}
where $T_{0}$ is the characteristic scale of the spin fluctuations. 
For a  review of experimental properties of these materials
see Stewart\cite{Ste}.

One of the central challenges of heavy fermion materials is to
understand the mechanism by which magnetism develops within the heavy
electron fluid.  Traditionally, magnetism and heavy fermion behavior
have been regarded as two mutually exclusive states, 
separated by a single quantum critical point. 
However, a variety of recent experiments suggest a richer state of
affairs, in particular: 
\begin{itemize}
\item [-] YbRh$_2$Si$_2$ can be driven to a quantum critical point by the
application of magnetic field, where both the N\'{e}el temperature and
the Kondo energy scale appear to simultaneously vanish. However, 
when doped, these two energy scales appear to separate from
one-another, indicating that the break-down of Fermi liquid behavior 
and the development of magnetism are not rigidly 
pinned together \cite{Pas,Geg2,Fri1};

\item [-]  In the 115 superconductor CeRhIn$_{5}$ there is evidence for a 
microscopic and homogeneous coexistence of local moment 
magnetism and superconductivity under pressure \cite{Kne};

\item [-] Neutron scattering experiments observe a partially ordered state in the geometrically frustrated CeP\nolinebreak dAl, in which one third of the Ce moments do not participate in the long-range order, suggesting the development of inhomogeneous Kondo states \cite{CePdAl1,CePdAl2, Fri2}.

\end{itemize}
%Phenomenology
Various phenomenological frameworks have been proposed for the
understanding of heavy fermion systems. The classical framework
proposed in the 70's by Doniach \cite{Don}, involves a 
competition between $T_K$ and $T_{RKKY}$ determining the ground state
to be a heavy Fermi liquid or magnetically ordered. More recently a
new axis was added to this picture, by the inclusion of geometric
frustration or reduction of dimensionality \cite{Col2, Si2}. These two
factors contribute towards the suppression of magnetism in a different
way, if compared to the competition with the Kondo effect. Also, based
on experiments in several families of heavy fermions, a
phenomenological two-fluid picture was proposed by Nakatsuji and
Pines, with predictive power on the ground state \cite{Nak,Yan}.

%Theory
Unfortunately these proposals do not give us information about the
character of the transition between the HFL and magnetic phases, and
its theoretical description has remained an unsolved challenge for several
decades. Theoretical proposals based on a spin density wave
description of the QCP \cite{Her, Mil, Mor}, Kondo breakdown
\cite{Col3}, deconfined quantum criticality \cite{Sen} and local
quantum criticality \cite{Si3} have been suggested, but no one picture
is yet able to fully account for experimental observations.

\subsection{Spin Representations in the Kondo Model}

The Kondo lattice Hamiltonian
\begin{eqnarray}
H_{KL}=\sum_{\bk\sigma} \epsilon_{\bk} c_{\bk\sigma}^\dagger c_{\bk\sigma}+J_K \sum_{i} \mathbf{S}_i \cdot \mathbf{s}_i,
\end{eqnarray}
provides a minimal model for heavy fermion systems.
The first term in $H_{KL}$ describes a band of conduction electrons with dispersion $\epsilon_\bk$, $J_K$ is the antiferromagnetic Kondo coupling between the local moment $\mathbf{S}_i$ and the spin of the conduction electron $\mathbf{s}_i$ at site $i$. 

%Why SUSY spins
The local moments  are neutral entities 
uniquely characterized by their spin quantum numbers. 
The removal of the charge degrees of
freedom from the Hilbert space of the localized f-electrons means that spin operators do not follow
canonical commutation relations; consequently, their treatment within a path
integral or diagrammatic 
approach is complicated by the absence of a Wick's theorem. 
To circumvent this difficulty, the spin operator is
traditionally factorized in terms of creation and annihilation operators:
\begin{eqnarray}
S_{\alpha\beta}=a_{\alpha}^\dagger  a_{\beta},
\end{eqnarray}
where $a_\alpha^\dagger$, $a_\alpha$ are bosonic or fermionic creation and annihilation operators, respectively,  and the indexes
$\alpha,\beta=\{1,2\}$ for an $SU(2)$ spin. There are actually several such
spin representations: the Holstein-Primakoff \cite{Hol}, Schwinger boson
\cite{Sch}, Abrikosov pseudo-fermion \cite{Abr} and the drone or
Majorana fermion
\cite{Ken} representations, among others. 

The physics that each of
these representations describes is profoundly different. 
For
example, 
the antiferromagnetic (AFM) phase at small 
$J_K$ is very effectively described by a
Schwinger boson representation of the local moments, with the
condensation of the bosons corresponding to the onset of magnetic
order \cite{Aue, Yos}. 
By contrast, the heavy Fermi liquid 
phase at large $J_{K}$
is successfully captured 
by a fermionic representation of the spins \cite{Col4}. 
We take the view that the success of these two representations in the different limits is
not simply one of mathematical convenience; rather, it 
reflects the physical transformation of both the spin
correlations and the excitations of the local moments: these evolve from
collective spin waves to charged heavy fermions.
Remarkably,
experiment indicates that these two phases connect together {\sl
continuously} via a quantum critical point, suggesting that at quantum
criticality the two representations merge. 

In this paper we argue that a full description of heavy fermion
materials requires a methodology that can capture
the transformation in the character of the ground-state and its
spin excitations.  This, in turn,
leads us to adopt a 
{\sl supersymmetric} representation of the spin\cite{Col1}
\begin{eqnarray}
S_{\alpha\beta}= f_{\alpha}^\dagger f_{\beta} +b_{\alpha}^\dagger
b_{\beta}.\end{eqnarray}
Here, $f_\alpha^\dagger$, $f_\alpha$ 
and $b_\alpha^\dagger$, $b_\alpha$ are respectively, fermionic and 
bosonic  creation and annihilation
operators. The spin is {\sl supersymmetric} because it is invariant under transformations that take bosons into fermions and vice versa; these are generated by fermionic operators which will be introduced in the next section.

\begin{figure}[here]
\begin{center}
\includegraphics[width=\columnwidth]{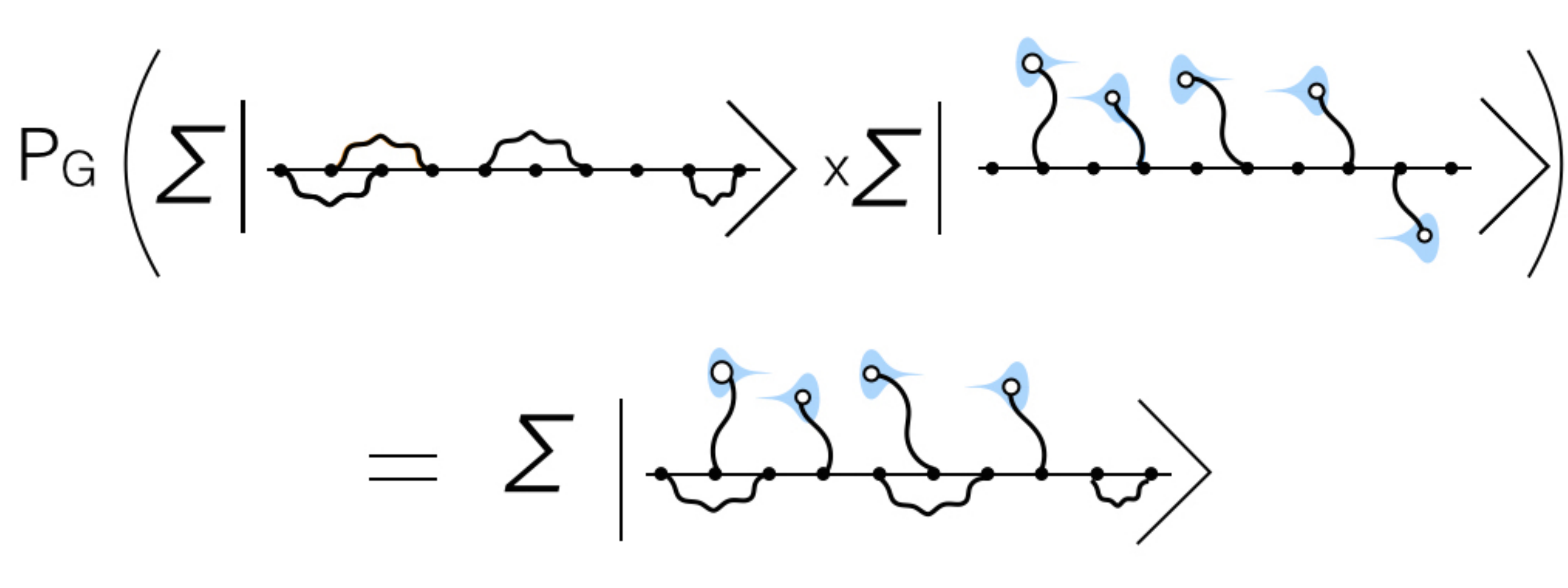}
\end{center}
\caption{Schematic illustration of a Gutzwiller wavefunction formed from
the projected product of a bosonic RVB wavefunction and a
Kondo-screened Fermi liquid, to form an entangled combination of both
wavefunctions. }
\label{fig:susypsi}
\end{figure}

One of the challenges of such a factorization, is that 
it requires a constraint which guarantees 
that the physics lies within the physical 
Hilbert space \cite{FPtrick}. For example, 
an elementary spin $S=1/2$  Kramers
doublet requires the constraint 
$Q=n_{b}+n_{f}=1$. 
Within this constrained Hilbert space, the most general wavefunction 
is an {\sl entangled product}
\begin{eqnarray}
|\Psi\rangle = P_G \bigl(\vert \Psi_F\rangle \otimes |\Psi_B\rangle\bigr),
\end{eqnarray}
where $\vert \Psi_{B}\rangle $ and $\vert \Psi_F\rangle$ are the bosonic and
fermionic components of the wavefunction, respectively,
while  $P_G$ is a Gutzwiller
projection operator. This operator can be written as:
\begin{eqnarray}\label{Gutz}
P_G =\int \Pi_i \frac{d \theta_i}{2\pi} e^{i\theta_i(n_{Bi}+n_{Fi}-1)}.
\end{eqnarray}
which imposes the constraint $n_{Bi}+n_{Fi}=1$ at each site $i$.
The unprojected wavefunction $\vert \psi_{B}\rangle $
describes the formation of long-range magnetic correlations in the
form of a bosonic RVB wavefunction, while $\vert \psi_{F}\rangle $
captures the development of Kondo of singlets 
and the development of a large Fermi surface of heavy
electrons.  The Gutzwiller projection entangles the two components of the
wavefunction into a single entity as illustrated
in Fig. \ref{fig:susypsi}.
%%%%%%%%%%%%%%Figure %%%%%%%%%%%%%%%%%%%%%%%%%%%%%%%%%%%%%%%%%%%%%%%%%%%
\vspace{1cm}
%%%%%%%%%%%%%%Figure %%%%%%%%%%%%%%%%%%%%%%%%%%%%%%%%%%%%%%%%%%%%%%%%%%%
This merged wavefunction has, in principle, the potential to capture the
two-fluid aspect of the heavy fermion ground-state. 

\begin{figure}[b]
\begin{center}
\includegraphics[width=0.4\linewidth, keepaspectratio]{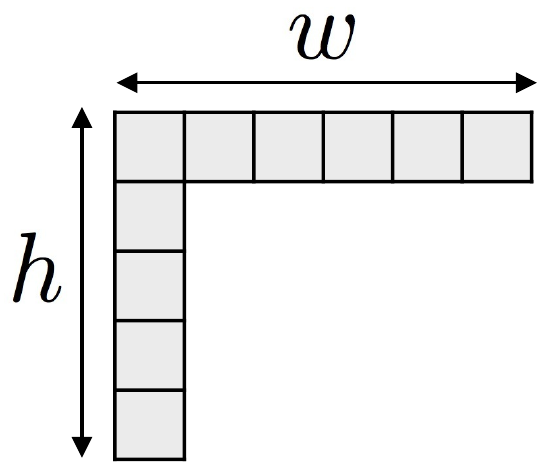}
\caption{Showing an L-shaped Young tableau,
characterized by two parameters, the width $w$ and the height $h$ of the tableau.}
\label{Lshaped}
\end{center}
\end{figure}

\subsection{Large-$N$ Approach}

The absence of a small parameter in the Kondo model effectively rules 
out the use of conventional 
perturbation theory. The alternative
approach, followed here, is the use of a large-$N$ expansion in which
the fundamental representations contain $N$, rather than $2$
components. In this approach $1/N\sim \hbar_{s}$ plays the role of synthetic
Planck's constant leading to a controlled mean-field (``classical'')
theory in the large-$N$ limit, with the possibility of
expanding the
fluctuations and the constraint condition as a power-series 
in $1/N$ about the large-$N$ limit. 
The simplest
generalization takes $SU(2)$ to
$SU(N)$\cite{readnewns,coleman,auerbach,Col6,millis,Aro}. Written in
traceless form the $SU (N)$ spin is then
\begin{eqnarray}
S^{SU(N)}_{\alpha\beta}&=& f_{\alpha}^\dagger f_{\beta}
+b_{\alpha}^\dagger b_{\beta }
- \left({n_{F}+n_{B}}
 \right)\delta_{\alpha \beta }/{N}
,
\end{eqnarray}
where  $\alpha, \beta\in \{1,2,...,N\}$. 
However, in this paper we seek to extend the supersymmetric
description of spins to the 
symplectic subgroup $SP(N)$ of $SU (N)$\cite{Rea1,Fli1,Fli2}:
\begin{eqnarray}
S^{SP(N)}_{\alpha\beta}&=& f_{\alpha}^\dagger f_{\beta}+b_{\alpha}^\dagger b_{\beta}-\tilde{\alpha}\tilde{\beta} (f_{-\beta}^\dagger f_{-\alpha}
+b_{-\beta}^\dagger b_{-\alpha}).
\end{eqnarray}
Here $N$ must be even, while the range of the elementary spin quantum
numbers is  $\alpha, \beta\in \{\pm 1,\pm 2,...,\pm N/2\}$. 
The tilde notation, employed extensively in this article,  denotes the sign of the index 
\begin{equation}\label{}
\tilde{\alpha }\equiv {\rm
sgn} (\alpha ),
\end{equation}
with the analogous definition for other indexes.
This new  spin operator 
has the symplectic property $S^{SP (N)}_{\alpha \beta } = - \tilde{\alpha
}\tilde{\beta }S^{SP (N)}_{-\beta ,-\alpha }$ (and is thus also traceless).
The symplectic group $SP (N)$ offers many
advantages for condensed matter physics, allowing for a consistent
extension of the notion of time reversal symmetry to the large-$N$
limit, which permits one to form singlet pairs of particles 
that are absent in the $SU (N)$ generalization \cite{Fli1,Fli2}. This capability is vital to describe
antiferromagnetism and superconductivity. 

The concept of a supersymmetric spin was introduced in previous
studies of impurity Kondo models \cite{Gan1,Gan2,Pep1,Col1}.  In the
work presented here, we follow the lines of Coleman et al.\cite{Col1},  with the
additional generalization to $SP(N)$, and discuss the spin representations in
terms of Young tableaux. Young tableaux provide a precise 
pictorial rendition of irreducible spin representations:
horizontal Young tableaux label
completely symmetric representations, which are naturally described by
bosons, while vertical Young tableaux label completely antisymmetric
representations, usually described by fermions. The use of 
supersymmetric representations lead us 
to consider the set of representations
characterized by L-shaped Young tableaux (Fig. \ref{Lshaped}). 
These
representations are characterized by two constants:

\begin{enumerate}
\item [-] the total number of elementary spins (or boxes) in the representation
$Q=h+w-1$, where $h$ and $w$ are height, and width of the Young
tableau, respectively, and
\item [-] 
the asymmetry $Y=h-w$ of the
L-shaped Young tableau, as discussed in Coleman et al.\cite{Col1}. 
\end{enumerate}

\begin{figure}[t]
\begin{center}
\includegraphics[width=0.9\linewidth, keepaspectratio]{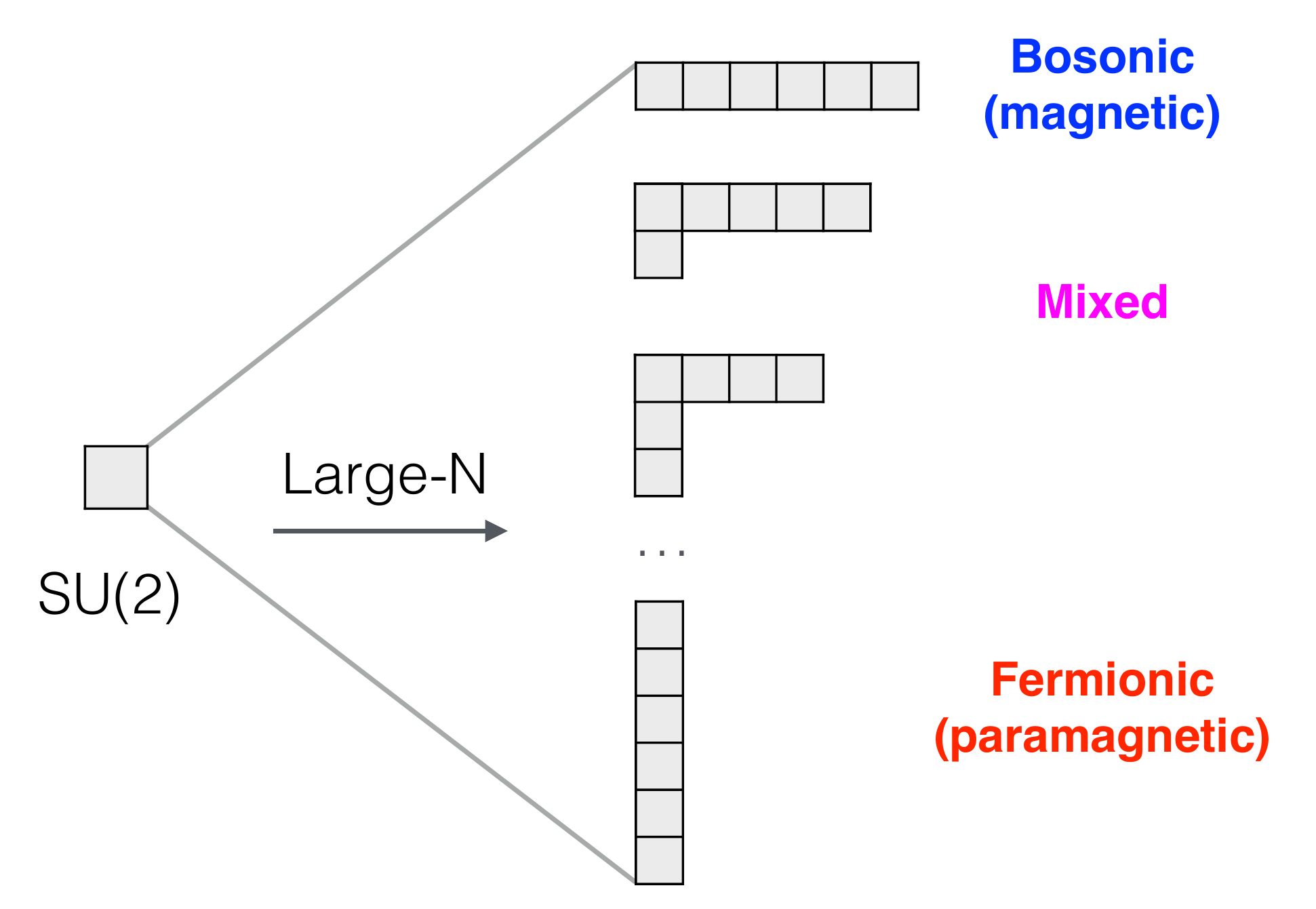}
\caption{Series of i tableaux in the large-$N$ limit ranging from a fully symmetric (top) towards a fully antisymmetric (bottom), passing through a series of L-shape representations.}
\label{Rep}
\end{center}
\end{figure}

The asymmetry of the representation 
is absent in a physical 
$SU(2)$ spin-1/2, in which case the Young tableau is depicted by a single box, but once we enlarge the
symmetry group of the spin in order to develop a large-$N$ theory,
we find a family of
representations that range from a completely symmetric representation,
fully described by bosons, to a completely antisymmetric
representation, described only by fermions, including a whole plethora
of intermediate representations that we refer to as \emph{mixed
representations}, depicted by L-shaped Young tableaux (see
Fig.~\ref{Rep}). The possibility of mean-field solutions described by
mixed representations is interesting as it may permit the description of new states
of matter, including coexistence of magnetism with superconductivity or with
 heavy Fermi liquid phases.

For a given value of $N$, one needs to decide which representation to
choose in order to proceed with the calculations. Traditionally a
purely bosonic representation or a purely fermionic
representation is chosen, but the supersymmetric approach provides the
possibility of considering an 
L-shaped representation. To constrain the problem to
such a representation one must fix the values of $\hat{Q}=Q_0$ and
$\hat{Y}=Y_0$ through the introduction of projection
operators into the partition function:
\begin{eqnarray}
Z=Tr[P_{Q_0,Y_0} e^{-\beta H}].
\end{eqnarray}

Typically, the more negative $Y$, the more symmetric 
the spin representation and the more
magnetic the resulting ground-state whereas the more positive $Y$, 
the more antisymmetric the spin representation and the more Fermi-liquid like the
ground-state. To avoid biasing the physics, we consider a
grand-canonical ensemble of representations defined by the partition
function with indefinite asymmetry $Y$, 
\begin{eqnarray}
Z=Tr[P_{Q_0} e^{-\beta H}]=\sum_{Y_0} Tr[P_{Q_0, Y_0} e^{-\beta H}],
\end{eqnarray}
where now we can identify $P_{Q_0}$ with the large-$N$ generalization of the Gutzwiller projection operator introduced in Eq.~\ref{Gutz}:
\begin{eqnarray}
P_G \Rightarrow P_{Q_0} =\int \Pi_i \frac{d \theta_i}{2\pi} e^{i\theta_i(n_{Bi}+n_{Fi}-Q_0)}.
\end{eqnarray}

This procedure will enable the ensemble to explore the lowest energy
configurations. 
Another motivation to work with the constraint that fixes only the
total number of boxes of the representation is the fact that the
asymmetry of the representation appears only in the large-$N$ limit, so
by letting $Y$ run free provides  an unbiased way 
to take the limit $N\rightarrow 2$, as schematically shown in Fig.~\ref{Rep}.

\subsection{New features of this work}

The large-$N$ limit we now develop places all L-shaped representations
with a given number of boxes on the same footing, and the asymmetry of the representation can be thought of as a
variational parameter. The character of the representation (bosonic,
fermionic or mixed) will now be decided by the energetics of the
problem. This will permit us 
to explore
the phase diagram of systems as heavy fermions, in which the character
of the spin changes from fermionic in the HFL phase towards bosonic in
the AFM region.

\begin{figure}[t]
\begin{center}
\includegraphics[width=\linewidth, keepaspectratio]{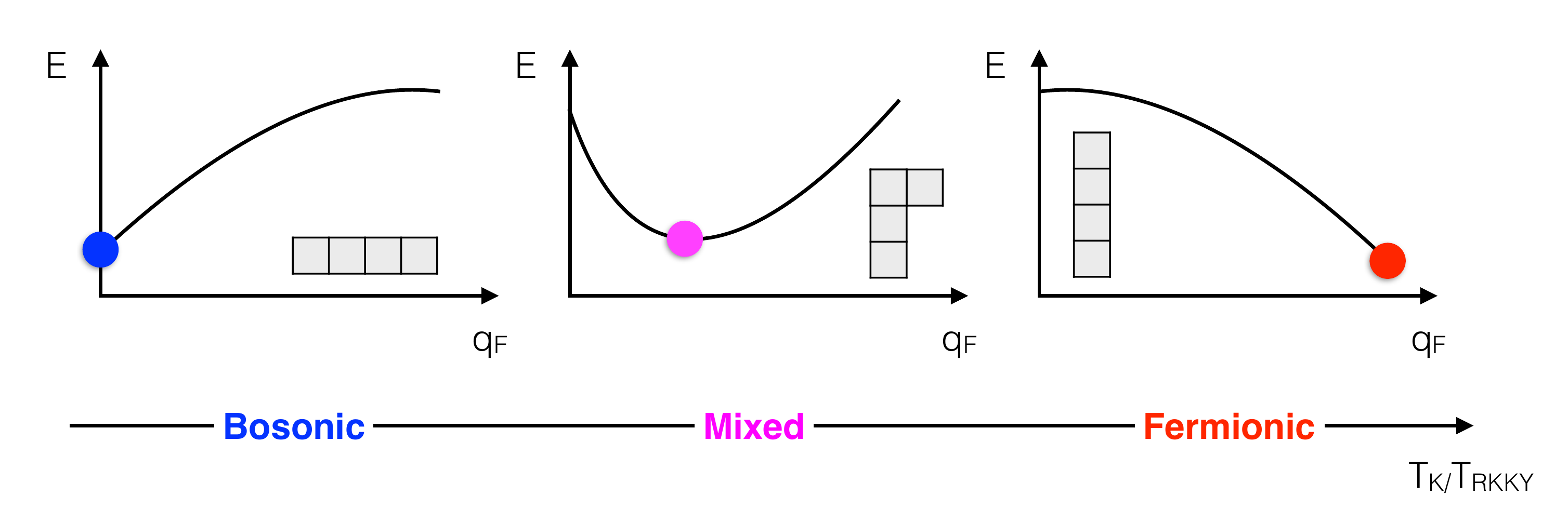}
\caption{Schematic representation of the evolution of the energy landscape as a function of $q_F$ for different values of the ratio $T_K/T_{RKKY}$.}
\label{SPoint}
\end{center}
\end{figure}

Fig. \ref{SPoint} illustrates schematically the evolution of the energy landscape (energy as a function of $q_{F}$, the number of fermions in the representation),  for three different values of  $T_K/T_{RKKY}$. For small $T_K/T_{RKKY}$ the energy landscape has a minima for $q_F=0$, which means that the system prefers to have a bosonic  spin representation and possibly develops magnetic order. Analogously, for large $T_K/T_{RKKY}$, the energy
landscape has a minimum for the maximum value of $q_F$, indicating a
purely fermionic representation, which would possibly lead to the development of
a heavy Fermi liquid. For intermediate values of $T_K/T_{RKKY}$ we
find that the representation is {\sl mixed}, with an energy minima
developing 
at an intermediate value of $q_F$ so that the minima is a saddle point as a function of
$q_{F}$. We are thus able to identify two classes of solution: 

\begin{enumerate}

\item[-] \underline{Type I minima}, in which the 
free energy is minimized by a purely bosonic or a purely fermionic
representation, indicating that the original supersymmetry of the
spin is severely broken. In this case the results of a purely bosonic
or fermionic representations are recovered;

\item[-] \underline{Type II minima}, in which 
mixed representations are energetically favorable. These kinds of minima
are candidate representations for a two-fluid picture 
of heavy fermions. Since the fermionic and bosonic components of the
spin fluid acquire the same chemical potential, this opens up the 
possibility of a new kind of zero mode: a \emph{Goldstino},
arising from the zero energy cost of rotations between the fermionic
and bosonic spin fluid. 

\end{enumerate} 

Other new features of this work are:

\begin{itemize}
\item [-] {\sl Symplectic Spins}: we generalize the rotation group of the
spin from $SU(2)$ to $SP(N)$ for a large-$N$ treatment
\cite{Fli1,Fli2}. This generalization guarantees that all the
components of the generalized spin properly invert under time reversal
which confers various advantages. In particular, it allows the description
of geometrically frustrated magnetism\cite{Rea1}
%one can use the same spin
%representation in every site of a lattice and be able to form singlets
%between those 
%
%, which is not possible within an $SU(N)$
%generalization, in which case one is restricted to treat only
%bipartite lattices with conjugate representations in different
%sub-lattices \cite{Aro}. 
and it permits the exploration of 
singlet superconductivity within the large-$N$ framework. 
%, since 
%one can now also decouple the
%four-body terms in the particle-particle channel;

%\item [-] Spin representation ensemble: in the infinite-N limit, we are not determining the spin representation ad hoc, but we are allowing the system to explore all the equivalent spin representations within a fixed number of boxes $Q_0$ and to select the one that minimizes the energy of the problem;
\end{itemize}

\begin{figure}[h]
\begin{center}
\includegraphics[width=0.75\linewidth, keepaspectratio]{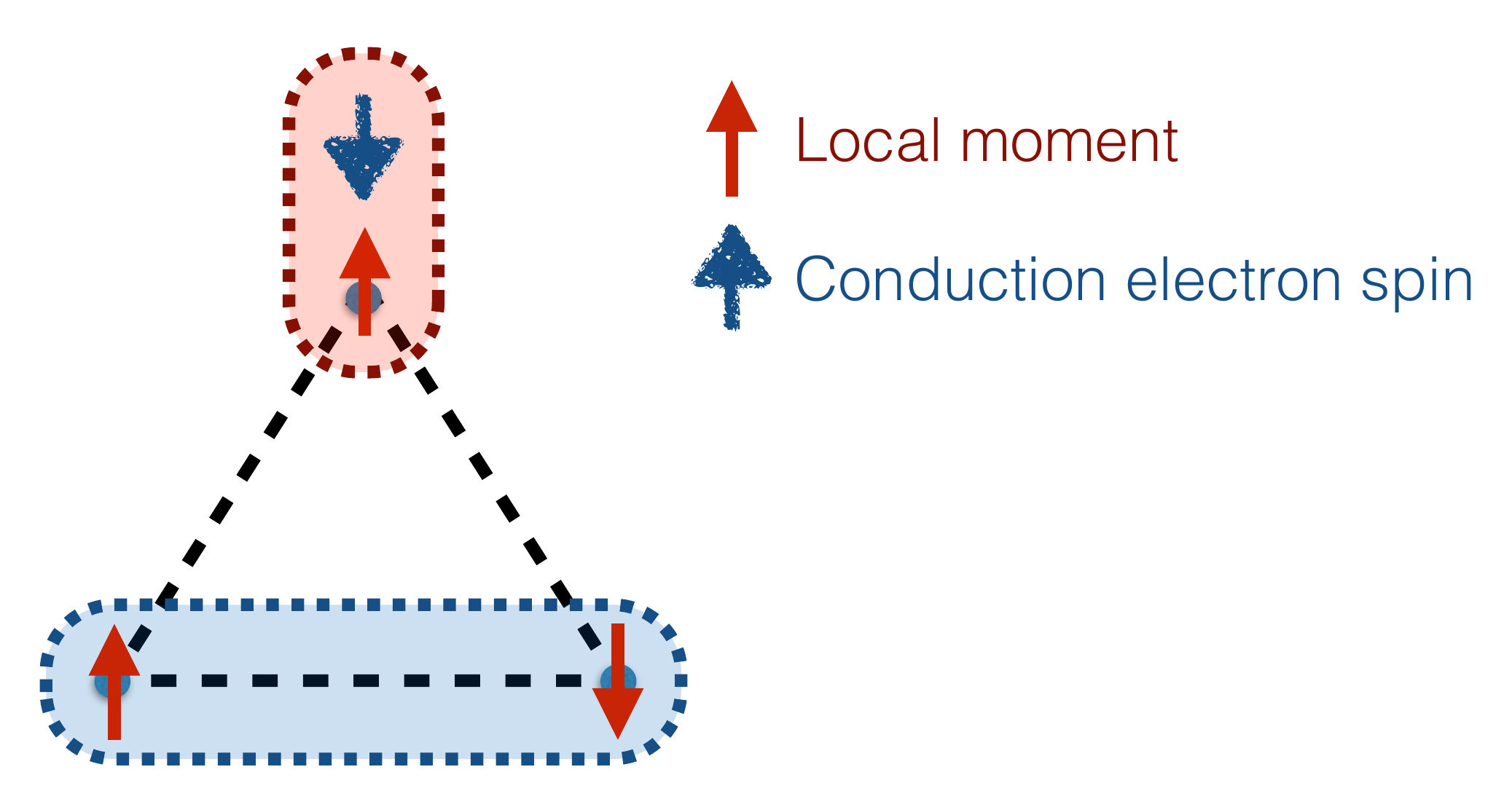}
\caption{Schematic representation of the inhomogeneous solution for the frustrated triangular geometry.}
\label{Inhom}
\end{center}
\end{figure}

\begin{itemize}
\item [-] {\sl Spatially inhomogeneous representations}: We explore the
possibility of solutions  which {\sl spontaneously} develop Kondo, 
or magnetic character 
at different sites. 
Here we are motivated by the partially ordered phase
verified experimentally in CePdAl. Fig.~\ref{Inhom} represents this
kind of solution schematically in a frustrated triangular geometry:
one of the local moments in the triangle develops fermionic character,
forming  a singlet with electrons in the conduction
sea, while the other two local moments have a bosonic representation,
forming an antiferromagnetic bond.

\end{itemize}

This paper is organized as follows:  in Section~\ref{SecSS} we define
the supersymmetric-symplectic spin, 
identifying the gauge group
under which it is invariant and discuss the Casimir for a given
irreducible representation. Details on the derivations are given in
Appendix~\ref{AppSS}. In Section~\ref{SecFor} we introduce the general
path integral formalism which leads to a mean field free energy in the
large-$N$ limit. We apply this formalism to a two-impurity model in
Section~\ref{Sec2Imp}, and details on the calculations are given in Appendices~\ref{AppFE}, \ref{AppBE} and \ref{AppFlu}. In Section~\ref{Sec3Imp} we apply the same formalism to a three-impurity model. We conclude and discuss the open questions in Section~\ref{SecCon}.

%%%%%%%%%%%%%%%%%%%%%%%%%%%%%%%%%%%%%%%%%%%%%%

\section{Supersymmetric-symplectic spins: definitions and properties}\label{SecSS}

%Definition
We start defining the supersymmetric-symplectic spin:
\begin{eqnarray}\label{S}
\hspace{-0.4cm}S_{\alpha\beta}= f_{\alpha}^\dagger f_{\beta} - \tilde{\alpha}\tilde{\beta} f_{-\beta}^\dagger f_{-\alpha} +b_{\alpha}^\dagger b_{\beta}  - \tilde{\alpha}\tilde{\beta} b_{-\beta}^\dagger b_{-\alpha},
\end{eqnarray}
where $f_\alpha^\dagger$, $f_\alpha$ 
and $b_\alpha^\dagger$, $b_\alpha$ are respectively, fermionic and 
bosonic  creation and annihilation
operators, with indexes $\alpha, \beta=\{\pm 1, \pm 2, ..., \pm N/2\}$, and $\tilde{\alpha}=sgn(\alpha)$,  $\tilde{\beta}=sgn(\beta)$. In this form the invertion of the spin under time reversal is made explicit. 

We can write the spin operator more concisely as:
\begin{eqnarray}\label{Ssusy}
S_{\alpha\beta}= \Psi^\dagger_{\alpha} \gamma_0 \Psi_{\beta} = \bar{\Psi}_{\alpha}\Psi_{\beta},
\end{eqnarray}
by introducing the four component spinor $\bar{\Psi}_\alpha=\Psi_\alpha^\dagger \gamma_0$, 
\begin{eqnarray}\label{Phi}
\Psi_{\alpha}=\left( \begin{array}{c}
f_\alpha\\
\tilde{\alpha} f_{-\alpha}^\dagger\\
b_\alpha\\
\tilde{\alpha} b_{-\alpha}^\dagger
\end{array} \right),
\end{eqnarray}
which carries the explicit spin index $\alpha$,  and has an implicit 
super-index which runs from 1 to 4 related to the supersymmetric and particle-hole character of its entries; and the matrix
\begin{eqnarray}
\gamma_0=\left( \begin{array}{cccc}
1 & 0 & 0 & 0\\
0 & 1& 0 & 0\\
0 & 0 & 1 & 0\\
0 & 0 & 0 & -1
\end{array} \right).
\end{eqnarray}
We shall follow the convention that super-indices are suppressed and
fully contracted with one-another  in our formulae, unless otherwise
stated. 

%Symmetry
The supersymmetric-symplectic spin defined in Eq.~\ref{Ssusy} commutes with the following operator bilinears and the respective conjugates:
\begin{align}
\hat{n}_F&=\sum_\alpha f_\alpha^\dagger f_\alpha,\nonumber\\\label{DefOp}
\hat{n}_B&=\sum_\alpha b_\alpha^\dagger b_{\alpha},\nonumber\\
\hat{\psi}&=\sum_{\alpha>0} f_\alpha f_{-\alpha},\\
\hat{\theta}&=\sum_\alpha  b_\alpha^\dagger f_\alpha,\nonumber\\
\hat{\eta}&=\sum_\alpha \tilde{\alpha}   f_{\alpha} b_{-\alpha} .\nonumber
\end{align}
These are therefore  generators of the symmetry group of the supersymmetric-symplectic spin. Note that $\hat{\psi}$ and $\hat{\eta}$ are present in the $SP(N)$ but not in the $SU(N)$ generalization of the supersymmetric spin \cite{Col1}. 

%The Super Group
We can rewrite these operators in the form of Hubbard operators \cite{Hubbard} as follows:
\begin{eqnarray}\label{thehubbard}
X_{00}&=&\hat{n}_B,\\
X_{++}&=&\frac{(\hat{n}_F-\hat{n}_B)}{2},\\
X_{--}&=&\frac{(N-\hat{n}_F-\hat{n}_B)}{2},\\
X_{+-}&=&\hat{\psi}^\dagger,\quad 
X_{-+}=\hat{\psi}
,\\
X_{+0}&=&\frac{\hat{\theta}^\dagger}{\sqrt{2}},\quad X_{0+}= \frac{\hat \theta}{\sqrt{2}}\\
X_{-0}&=&\frac{\hat{\eta}}{\sqrt{2}},\quad X_{0-}=\frac{ \hat{\eta }\dg }{\sqrt{2}},
\end{eqnarray}
where $X_{00}$, $X_{\pm \pm }$ and $X_{\pm \mp}$
are bosonic Hubbard operators, while $X_{\pm 0}$ and $X_{0\pm}$ are fermionic Hubbard operators. In this form the algebra that these operators follow can be concisely written as:
\begin{eqnarray}\label{hubbardalgebra}
[X_{ab},X_{cd}]_{\pm}=X_{ad}\delta_{bc}\pm X_{cb}\delta_{ad},
\end{eqnarray}
where the anticommutator, $(+)$, is used when both Hubbard
operators are fermionic, while the commutator, $(-)$, otherwise. The
Hubbard algebra\cite{Hubbard} above defines the $SU (2|1)$
supergroup\cite{Bar}. One can explicitly see the $SU (2)$ subgroup
generated by the isospin operators 
\begin{eqnarray}
\Psi_1 &=& (X_{+-}+X_{-+})/2,\\
\Psi_2 &=& -i(X_{+-}-X_{-+})/2,\\
\Psi_3 &=& (X_{++}-X_{--})/2,
\end{eqnarray}
which follow the commutation relation:
\begin{eqnarray}
[\Psi_i,\Psi_j]=i\epsilon_{ijk}\Psi_k,
\end{eqnarray}
and are related to the rotation of the fermionic components of $\Psi_\alpha$; while $X_{00}$ defines the generator for the $U (1)$ subgroup associated with the bosons.  

Note that one can perform a super-rotation $g$ taking $\Psi_\alpha \rightarrow g \Psi_\alpha$ which leaves the spin invariant:
\begin{eqnarray}
S_{\alpha\beta}= \Psi^\dagger_{\alpha} \gamma_0 \Psi_{\beta} \rightarrow \Psi^\dagger_{\alpha} g^\dagger \gamma_0  g \Psi_{\beta}
\end{eqnarray}
if $g$ satisfies
\begin{eqnarray}
g^\dagger \gamma_0 g=\gamma_0.
\end{eqnarray} 

The most general transformation $g$ can be constructed by
exponentiation of the generators of the group listed above. An
explicit form for g is given in Appendix~\ref{AppA2}.

\subsection{The Casimir} To uniquely characterize an irreducible
representation of $SP(N)$, in principle one needs to define $r=N/2-1$
Casimirs, where $r$ is the rank of the group (the dimension of the
Cartan sub-algebra). In terms of Young tableaux, one can understand
these parameters as the number of boxes in each row of the tableau
(the maximum number of rows in the Young tableau for $SP(N)$ is
$N/2$). In this work we restrict our attention to L-shaped tableaux, so we
have the extra information that all the rows below the first one have
no more than a single box. This reduces the number of parameters
required to define the representation to two: $Q$, the total number of
boxes, and $Y$, the asymmetry of the representation, as discussed in
Coleman et al.\cite{Col1}.

From the second Casimir we can identify the quantities $Q$ and $Y$. From Nwachuku\cite{Nwa}, we can deduce that for an L-shaped representation in $SP(N)$ the second Casimir can be written in terms of the width $w$ of the first row and the height $h$ of the column in the tableau as:
\begin{eqnarray}\label{NwaEq}
C_2= 2(w+h)(N+w-h) + 4(h-N/2) -2,
\end{eqnarray}
and identifying $Q=w+h-1$ and $Y=h-w$, we have
\begin{eqnarray}\label{NwaEq2}
C_2= 2Q(N+1-Y),
\end{eqnarray}
the details of this derivation are shown in Appendix~\ref{AppSSNwa}.

In terms of operators, the second Casimir can be written as the
magnitude of 
the 
spin:
\begin{eqnarray}\label{S2}
\mathbf{S}^2 &=& \sum_{\alpha\beta} S_{\alpha\beta}S_{\beta\alpha}\\
&=&2\hat{Q}(N+1-\hat{Y})\nonumber
\end{eqnarray}
where:
\begin{eqnarray}
\hat{Q}&=&\hat{n}_F+\hat{n}_B,\\ 
\hat{Y}&=&\hat{n}_F-\hat{n}_B+1+\frac{4\hat{\psi}^\dagger \hat{\psi}-2 \hat{\theta}\dg  \hat{\theta} +2\hat{\eta}^\dagger \hat{\eta}}{\hat{Q}}.
\end{eqnarray}
This form is similar to that found for the $SU(N)$ case (see
Appendix~\ref{AppA4} for details of the derivation). Note that in
the lowest weight state $\vert \Psi_{0}\rangle $, where 
$\psi \vert  \Psi_{0}\rangle =\eta \vert  \Psi_{0}\rangle = \theta
\vert  \Psi_{0}\rangle =0$, (corresponding to no pairs and to
minimizing the number of fermions) $\mathbf{S}^{2}=
2 (n_{B}+n_{F}) (N -n_{F}+ n_{B})$. 

We can also write $\mathbf{S}^2$ in terms of the
$SU (2|1)$ Casimir of the Hubbard operators via the identity:
\begin{eqnarray}
(N^{2}-\mathbf{S}^2)/4=
X_{\alpha\beta}X_{\beta\alpha}-
[X_{\alpha 0},X_{0\alpha}]- (X_{00})^{2},
\end{eqnarray}
with an implied summation over the repeated indices $\alpha, \beta =
\pm $ (See Appendix~\ref{AppA5}).

%%%%%%%%%%%%%%%%%%%%%%%%%%%%%%%%%%%%%%%%%%%%%%

\section{The formalism}\label{SecFor}

The main topic of interest in our work is the class of Kondo-Heisenberg models involving $SP (N)$
 spins, interacting via 
an additional nearest neighbor antiferromagnetic 
Heisenberg exchange.  These  models are written as
\begin{equation}\label{}
H = H_{c}+ \sum_{j}H_{K} (j)  + \sum_{(i,j)}H_{H} (i,j)
\end{equation}
where
\begin{equation}\label{}
H_{c} = \sum_{\bk \alpha }\epsilon_{\bk }c\dg_{\bk \alpha }c_{\bk \alpha }
\end{equation}
describes a conduction band of electrons of dispersion $\epsilon_{\bk
}$, where $c\dg_{\bk \alpha }$ creates a conduction electron of
momentum $\bk $, spin index
$\alpha \in
\{\pm1,\pm2,...,\pm N/2\}$. The
term 
\begin{equation}\label{}
H_{K} (j) = \frac{J_{K}}{N}\sum_{\alpha \beta } S_{\alpha \beta } (j)s_{\beta \alpha } (j)
\end{equation}
describes the Kondo interaction at site $j$, where
$S_{\alpha\beta} (j)= \bar{\Psi}_{j\alpha}\Psi_{j\beta}$
defines the local moment operator as in Eq.~\ref{Ssusy} and the
conduction electron spin operators are also written in symplectic form:
\begin{eqnarray}
s_{\alpha\beta} (j)= 
c_{j\alpha}\dg c_{j\beta} -\tilde{\alpha}\tilde{\beta}c_{j-\beta}\dg c_{j-\alpha}.
\end{eqnarray}
The final term describes the Heisenberg interaction between spins at
sites $i$ and $j$, given by
\begin{equation}\label{}
H_{H} (i,j) = \frac{J_{H}}{N} \sum_{\alpha \beta }  S_{\alpha \beta } (i)S_{\beta \alpha } (j).
\end{equation}
To display the supersymmetric gauge character of the interactions we
now rearrange the order of the operators.  First, using the properties
of the symplectic spin, 
$\tilde{\alpha }\tilde{\beta }S_{\alpha \beta }= -S_{-\beta,
-\alpha }$, the two parts of the electron spin operator in the Kondo interaction are folded
into one as follows: 
\begin{eqnarray}\label{l}
H_{K} (j) &=& 
\frac{J_{K}}{N}\sum_{\alpha\beta }S_{\alpha \beta } (j)
(c_{j\beta}\dg c_{j\alpha } -\tilde{\alpha}\tilde{\beta}c_{j-\alpha}\dg c_{j-\beta}
)\cr
&=&
\frac{2J_{K}}{N}\sum_{\alpha\beta }\bar  \Psi_{\alpha } (j) \Psi_{\beta } (j)c_{j\beta}\dg c_{j\alpha }.
\end{eqnarray}
Next, we super-commute the $\bar \Psi  $ field to the right-hand
side of the interaction, rewriting the interaction in terms of a
supertrace, 
\begin{equation}\label{}
H_{K} (j) 
= - \frac{2J_{K}}{N} \sum_{\alpha\beta }
{\rm Str}\left[\left(\Psi_{j\beta }c\dg_{j\beta } \right)
\left(c_{j\alpha}\bar \Psi_{j\alpha } \right) 
\right]
\end{equation}
where we define the supertrace as ${\rm
Str}[A]=A_{11}+A_{22}-A_{33}-A_{44}$ and we have 
used the property that the
dot product of two super-spinors $\bar \Phi $ and $\chi $ 
can be rewritten as a supertrace of their outer-product
$[\bar \Phi \chi ] = - {\rm Str}[\chi \bar \Phi ]$.
Notice that ${\nu}_{j}=\left(\Psi_{j\alpha }c\dg_{j\alpha } \right)
$
and  $\bar  {\nu}_{j}=\left(c_{j\beta}\bar \Psi_{j\beta } \right) 
$ are four component column and row spinors that respectively
transform like $\Psi_{j\alpha}$ and $\bar \Psi_{j\alpha}$ under super-rotations.

In a similar fashion, we rewrite the Heisenberg interaction as 
\begin{eqnarray}\label{l}
H_{H} (i,j) &=& \frac{J_{H}}{N}\sum_{\alpha\beta } 
\bar \Psi_{i\alpha} \Psi_{i\beta } \bar \Psi_{j\beta } \Psi_{j\alpha }
\cr
&=&
-\frac{J_{H}}{N}\sum_{\alpha\beta } 
{\rm Str}\left[
\left(\Psi_{i\beta }  \bar \Psi_{j\beta}  \right)
\left(\Psi_{j\alpha}
\bar  \Psi_{i\alpha } \right)
\right].
\end{eqnarray}
Notice  that the object
$u_{ij}=\left(\Psi_{i\beta }  \bar \Psi_{j\beta}  \right)
$ is an outer-product of two super-spinors, forming a four-by-four
tensor in superspace that transforms as $u_{ij}\rightarrow
g_{i}u_{ij}g_{j}^{-1}$ under super-rotations.

\begin{widetext}
With these manipulations, the Kondo-Heisenberg model
can be written as:
\begin{eqnarray}
H=H_c
-\frac{2J_{K}}{N}\sum_{j,\alpha \beta}{\rm Str}\left[
\left(\Psi_{j\alpha }c\dg_{j\alpha } \right)
\left(c_{j\beta}\bar \Psi_{j\beta } \right) \right]
-\frac{J_{H}}{N}\sum_{(i,j)\alpha ,\beta } 
{\rm Str}\left[
\left(\Psi_{i\beta }  \bar \Psi_{j\beta}  \right)
\left(\Psi_{j\alpha}
\bar  \Psi_{i\alpha } \right)
\right].
\end{eqnarray}
Notice that 
the invariance property  of the supertrace ${\rm Str}[g T g^{-1}]=
{\rm Str}[T]$ guarantees that these interactions are gauge-invariant
under the local super-rotations $\Psi_{j}\rightarrow
g_{j}\Psi_{j}$.
The factorized forms of the interactions are convenient for
Hubbard-Stratonovich transformations. 
\end{widetext}

The constraint fixing the total number of bosons plus fermions at each site $n_{Fi}+n_{Bi}=Q_0$ can also be written in terms of the spinors $\Psi_{i\alpha}$:
\begin{eqnarray}
n_{Fi}+n_{Bi}=\frac{1}{2}\sum_\alpha \bar{\Psi}_{i\alpha} \Lambda \Psi_{i\alpha},
\end{eqnarray}
where
\begin{eqnarray}
\Lambda=\left( \begin{array}{cccc}
1& 0 & 0 & 0\\
0 & -1& 0 & 0\\
0 & 0 & 1 & 0\\
0 & 0 & 0 & -1
\end{array} \right).
\end{eqnarray}

We can now write the partition function as a functional integral over
the constraint field $\lambda$ and the spin carrying boson and
fermion fields
\begin{eqnarray}
Z=\int \mathcal{D}\lambda\mathcal{D}\mu e^{-S},
\end{eqnarray}
where 
\begin{eqnarray}
S= S_c+S_{S}+S_K+S_H,
\end{eqnarray}
while 
\begin{eqnarray}\label{measure1}
\mathcal{D}\mu=\mathcal{D}[ c, f, b], 
\end{eqnarray}
is the measure of integration over the canonical 
$c$, $f$ and $b$ fields and
\begin{equation}\label{}
{\cal D}\lambda =\prod_j d\lambda_{j}
\end{equation}
is the measure of integration over the constraint.

The components of the action are:
\begin{eqnarray}
S_c=\int_0^\beta d\tau \sum_{\bk\alpha} c^\dagger_{\bk\alpha}(\partial_\tau +\epsilon_{\bk})c_{\bk\alpha},
\end{eqnarray}
the conduction electron part of the action;
\begin{eqnarray}
S_{S}=\!\!\int_0^\beta\! \!d\tau \left[\frac{1}{2}\sum_{j,\alpha}
\bar{\Psi}_{j\alpha}(\partial_\tau+ \lambda_j \Lambda) \Psi_{j\alpha} 
-\sum_{j}\lambda_j Q_0\right]
\end{eqnarray}
describes the Berry phase, with the constraint $n_{Bj}+n_{Fj}=Q_0$
imposed via 
the introduction of the Lagrange multiplier $\lambda_j$ at
each site, 
while 
\begin{eqnarray}
S_{K}&=&
 \int_{0}^{\beta }d\tau 
\sum_{j}H_{K} (j), \cr  S_{H} &=& 
 \int_{0}^{\beta }d\tau \sum_{(ij) }H_{H} (i,j)
\end{eqnarray}
are the Kondo and Heisenberg parts of the action.

\begin{widetext}
Inside the path integral, we can now carry out a Hubbard-Stratonovich
transformation of the interactions. The Kondo part of the interaction
is factorized as follows
\begin{eqnarray}\label{l}
H_{K} (j)&=& - \frac{2J_{K}}{N} \sum_{\alpha\beta }
{\rm Str}\left[\left(\Psi_{j\alpha }c\dg_{j\alpha } \right)
\left(c_{j\beta}\bar \Psi_{j\beta } \right) 
\right] \rightarrow \cr
H_{K}' (j)&=& \sum_\alpha {\rm Str}\left[
\left(\Psi_{j\alpha }c\dg_{j\alpha
} \right)\bar V_{j}+ V_{j}\left(c_{j\alpha}\bar \Psi_{j\alpha } \right) 
+ \frac{N}{2J_{K}} \bar {V}_{j}V_{j} \right]\cr
&=&  
\sum_\alpha \left[
\left(\bar V_{j}\Psi_{j\alpha } \right) c\dg_{j\alpha}
+
c_{j\alpha} \left(
\bar{\Psi}_{j\alpha}V_{j}\right) 
\right] + \frac{ N}{2 J_K} Tr[ V_j \bar V_{j}].
\end{eqnarray}
In the last step we have absorbed the minus sign associated with the
anticommutation of $c\dg _{j\alpha }$ and $\left(\bar
V_{j}\Psi_{j\alpha } \right)$, likewise
$c_{j\alpha} $ and  $\left(
\bar{\Psi}_{j\alpha}V_{j}\right)$. These Hubbard-Stratonovich fields $V_{j}$ and $\bar V_{j}=
V\dg_{j}\gamma_0$ are four-component spinors 
\begin{eqnarray}
V_{j}=\left( \begin{array}{c}
v_{j} \\
{d}_{j} \\
 \phi_{j}  \\
{\xi}_{j} \end{array} \right), \qquad 
\bar  V_{j} = (\bar v_{j},\bar d_{j}, \bar \phi_{j},-\bar \xi_{j}).
\end{eqnarray}
Here $v_i$ and $d_i$ are complex fields related to the hybridization
between f-fermions and c-electrons and the development of
superconductivity by the formation of pairs between f-fermions and
c-electrons, respectively. The parameters $\phi_i$ and $\xi_i$ are
complex Grassmann numbers, the first related to the hybridization
between b-bosons and c-electrons and the second related to the
development os pairs formed between b-bosons and c-fermions.

In a similar fashion, the Heisenberg term decouples as:
\begin{eqnarray}\label{Str}
H_{H} (i,j) \rightarrow 
H_{H}' (i,j) &=&-\sum_{\alpha }{\rm Str}\left[ \Delta_{ij}\left(\Psi_{j\alpha}\bar  \Psi_{i\alpha } \right)
+ \left(\Psi_{i\alpha }  \bar \Psi_{j\alpha}  \right)
\bar \Delta_{ij}
\right]+ \frac{N}{J_{H}}
{\rm Str}\left[
\bar\Delta_{ij} \Delta_{ij}
 \right]\cr
&=& \sum_{\alpha }\left[ \bar \Psi_{i\alpha }\Delta_{ij}\Psi_{j\alpha }+
\bar \Psi_{j\alpha }\bar \Delta_{ij}\Psi_{i\alpha }
 \right] + \frac{N}{J_{H}}{\rm Str}\left[\bar  \Delta_{ij}\Delta_{ij}\right]
\end{eqnarray}
where $\Delta_{ij}$ is a
four-by-four matrix and 
its conjugate is defined as 
$\bar \Delta_{ij}= \gamma_{0}\Delta_{ij}\dg \gamma_0$. The structure
of the matrix is as follows
\end{widetext}
\begin{eqnarray}
\Delta_{ij}=\left( \begin{array}{cc}
\Delta_{F} & \tilde{\Delta}_{S} \\
{\Delta}_{S}& \Delta_{B} 
 \end{array} \right)_{ij},
\end{eqnarray}
where the diagonal block-matrices composed by c-numbers whereas the 
off-diagonal block matrices $\Delta_{S}$ and  $\tilde{\Delta
}_{S}$ are composed by Grassmanians. The internal structure of these blocks is
given by
\begin{eqnarray}
(\Delta_{F})_{ij}=\left( \begin{array}{cc}
t_{ij} & p_{ij}  \\
\bar{p}_{ij} & -\bar{t}_{ij}
 \end{array} \right),
  \end{eqnarray}
 \begin{eqnarray}
(\Delta_{B})_{ij}=\left( \begin{array}{cc}
q_{ij} & g_{ij} \\
 \bar{g}_{ij} & -\qb_{ij} \end{array} \right),
 \end{eqnarray}
 \begin{eqnarray}
 (\Delta_{S})_{ij}=\left( \begin{array}{cc}
\gamma_{ij}&\mu_{ij}  \\
-\bar{\mu}_{ij} & \bar{\gamma}_{ij} \\
\end{array} \right),
\end{eqnarray}
where the remaining matrix is defined through the relation
$(\tilde{\Delta}_{S})_{ij}= \sigma_{3}
(\Delta_{S})_{ji}$.
The matrix $\Delta_{ij}$ can be thought of as supersymmetric RVB
field. 
The components fields $t_{ij}$ and $p_{ij}$ promote hopping and
pairing amongst the f-fermions in different sites and the complex
fields $q_{ij}$ and $g_{ij}$ promote hopping and magnetic bond
formation between the b-bosons. The  Grassmannian parameters
$\gamma_{ij}$ are hopping amplitudes
that transmute 
bosons into fermions and vice versa, while the Grassmannian amplitudes
$\mu_{ij}$ describe pairing between 
bosons and fermions at different sites.

The partition function now reads:
\begin{eqnarray}\label{DDV}
&&Z=\int {\cal D}[\lambda ,V ,\Delta]{\cal D}\mu e^{-S'},
\end{eqnarray}
where 
\begin{eqnarray}
S'= S_c+S_{S}+S'_K+S'_H,
\end{eqnarray}
where the primes denote the actions of the Hubbard-Stratonovich factorized
interactions
\begin{eqnarray}\label{l}
S'_{K} &=& \int_{0}^{\beta }d\tau \sum_{j} H'_{K} (j),\cr
S'_{J} &=& \int_{0}^{\beta }d\tau  \sum_{\langle  i,j\rangle } H'_{H}
(i,j),
\end{eqnarray}
and the integral over $\mathcal{D}[V,\Delta]$ indicates the integral
over all the fluctuating fields introduced by the Hubbard-Stratonovich
transformations.  Here c-number fields are represented by the latin
letters $(v, d, p, t , q, g)$, while the Grassmannian fields are
represented by the Greek letters $(\phi,
\xi,\gamma,\mu)$. Grassmannian fields are introduced in order to
decouple terms with fermionic bilinears. Note that the Kondo and
Heisenberg parts of the action are invariant under the transformation
$\Psi_{j\sigma}\rightarrow g_{j} \Psi_{{j}\sigma}$ if the fluctuating field
matrices transform accordingly as 
\begin{eqnarray}
V_j&\rightarrow& g_{j} V_{j}
\end{eqnarray}
\begin{eqnarray}
\Delta_{ij}&\rightarrow&g_{i}\Delta_{ij} g_{j}^{-1}.
\end{eqnarray}

Now we move to the discussion of the implementation of the constraint
by fixing $\hat{Q}=Q_0$. The constraint can be imposed as a projection
operator in each site $j$, written as a delta function:
\begin{eqnarray}
P_{Q_0} = \Pi_j P_{Q_0}^j=\Pi_j \delta(\hat{Q}_j-Q_0).
\end{eqnarray}

In order to treat the bosonic and fermionic components of the spin in
the grand canonical ensemble, we split the constraint into two terms as follows:
\begin{eqnarray}
P_{Q_0}^j=\sum_{Q_{Fj}=0}^{Q_0}\delta(\hat{n}_{Fj}-Q_{Fj})\delta(\hat{n}_{Bj}-Q_{0}+Q_{Fj}).
\end{eqnarray}
The constraint fixes the total number of bosons and fermions to
$Q_0$, leaving the asymmetry of the representation free to adjust 
according to the energetics of the problem.

\begin{widetext}
This constraint can be implemented in the path integral as a Dirac delta function in its integral form:
\begin{eqnarray}\label{Constr}
P_{Q_0}&=& \int \mathcal{D}[Q_{F},\lambda] e^{-\sum_{j}S_P (j)}, \cr
\mathcal{D}[Q_{F},\lambda]&=&
\prod_{j} \sum_{Q_{Fj}}d\lambda_{Fj} d\lambda_{Bj},
\end{eqnarray}
where
\begin{eqnarray}\label{l}
S_P (j)=\int_0^\beta d\tau \biggl[ \lambda_{Fj}(n_{Fj}-Q_{Fj})
+\lambda_{Bj}(n_{Bj}-Q_{0}+Q_{Fj})\biggr],
\end{eqnarray}
and the constraint fields, $\lambda_{Fj}$ and  $\lambda_{Bj}$ 
are integrated along the imaginary axis. 
From these considerations $S_S$ can be rewritten as:
\begin{eqnarray}
S_{S}\rightarrow S'_S=\!\!\int_0^\beta\! \!d\tau\sum_j\! \Bigg(\sum_{\sigma} \bar{\Psi}_{j\sigma}\frac{(\partial_\tau+ \Lambda'_j)}{2} \Psi_{j\sigma}- \lambda_{Fj}Q_{Fj}-\lambda_{Bj} (Q_{0}-Q_{Fj}) \Bigg),
\end{eqnarray}
where now
\begin{eqnarray}
\Lambda'_j=\left( \begin{array}{cccc}
\lambda_{Fj} & 0 & 0 & 0\\
0 & -\lambda_{Fj}& 0 & 0\\
0 & 0 & \lambda_{Bj} & 0\\
0 & 0 & 0 & -\lambda_{Bj}
\end{array} \right),
\end{eqnarray}
and the partition function is now written as:
\begin{eqnarray}\label{DDV}
&&Z=\int \mathcal{D}[Q_{F},\lambda,V,\Delta ]
{\cal D}\mu e^{-S'},
\end{eqnarray}
where 
\begin{eqnarray}
S'= S_c+S'_{S}+S'_K+S'_H.
\end{eqnarray}
\end{widetext}
The new feature of this action, is the appearance of the term $Q_{F}$,
which tunes the bosonic/fermionic character of the representation. In
the large $N$ limit, we will be able 
to replace the discrete measure
of integration over $Q_{F}$ by a continuous measure
\begin{equation}\label{}
\sum_{Q_{Fj}}\rightarrow \int_{0}^{Q_{0}} dQ_{Fj}. \qquad  (N\rightarrow \infty ).
\end{equation}

In the large $N$ limit, we anticipate that 
the functional integral is given by the saddle point value of the
effective action, so the large $N$ approximation is then given by the
exponential of the effective action
\begin{equation}\label{}
Z\approx  
e^{-S_{eff}[Q_{F},\lambda,V,\Delta] }
\end{equation}
where
\begin{equation}\label{}
e^{-S_{eff}}= \int {\cal D}[c,f,b] e^{-S'}
\end{equation}
is the integral over the canonical spin-carrying fermions and bosons
in the presence of fixed $Q_{F}$ $\lambda_{B}$, $\lambda_{F}$, $V$ and $\Delta $
with the conditions that $S_{eff}$ be stationary with respect to each
of its fields. We may anticipate two classes of mean-field solution
\begin{enumerate}
\item Type I solutions, in which the minimum of the effective action
occurs at the extremum of the summation over $Q_{Fj}$, i.e
$Q_{Fj}=Q_{0}$ or $Q_{Fj}=0$, corresponding to the fully fermionic or
bosonic solutions.

\item Type II solutions, in which the minimum of the effective action
occurs at some intermediate value of $0<Q_{Fj}<Q_{0}$.  Since the
action is stationary with respect to variations in $Q_{Fj}$, this
implies that 
\begin{equation}\label{}
\frac{\delta S_{eff}}{\delta Q_{Fj}}= 0 = \lambda_{Bj}-\lambda_{Fj}
\end{equation}
so that in this phase, the chemical potential of the bosonic  and
fermionic spinons are equal, 
\begin{equation}\label{}
\lambda_{Bj}=\lambda_{Fj}.
\end{equation}
This equality of chemical potentials allows us to consider these
solutions as {\sl two fluid} solutions. 

\end{enumerate}
At the saddle points, we can set all fermionic components of $V$ and
$\Delta $ to zero. These terms only contribute to the fluctuations
about the mean field theory, so we have 
\begin{eqnarray}
V_{j}\rightarrow V_j^0=\left( \begin{array}{c}
v_{j} \\
d_{j} \\
0  \\
0 \end{array} \right),
\end{eqnarray}
and
\begin{eqnarray}
\Delta_{ij}\rightarrow\Delta_{ij}^0=\left( \begin{array}{cc}
\Delta_{F} &0\\
0& \Delta_{B} 
 \end{array} \right)_{ij}.
\end{eqnarray}

Note that the fermionic and bosonic parts of the action decouple since
all the matrices in the action now have the blocks linking the fermionic and bosonic subspaces  equal to zero. Now it is possible to solve the fermionic and bosonic
problems separately, imposing the constraint $Q_{Fj}+Q_{Bj}=Q_0$ to
the solution in the end of the calculation. Note that this provides a
picture of two asymptotically independent fluids, bosonic and
fermionic, in the large-$N$ limit and that the introduction of
fluctuations will provide interactions between them.

Now, as a first exploration of this idea we illustrate the formalism with two simple examples: a two-impurity and a frustrated three-impurity model and show that there are stable mean field solutions with mixed representations, as well as with purely bosonic and purely fermionic representations.

%%%%%%%%%%%%%%%%%%%%%%%%%%%%%%%%%%%%%%%%%%%%%%
\section{The  Two impurity model}\label{Sec2Imp}

%The model %Writing in terms of B and F
As a first application of the supersymmetric-symplectic spin, we study a minimal model that allows one to make connections to the physics of heavy fermion systems. The model consists of two local moments interacting among themselves by a Heisenberg coupling $J_H$ and interacting with its respective bath of conduction electrons by a Kondo coupling $J_K$.

\begin{figure}[h]
\begin{center}
\includegraphics[width=0.8\linewidth, keepaspectratio]{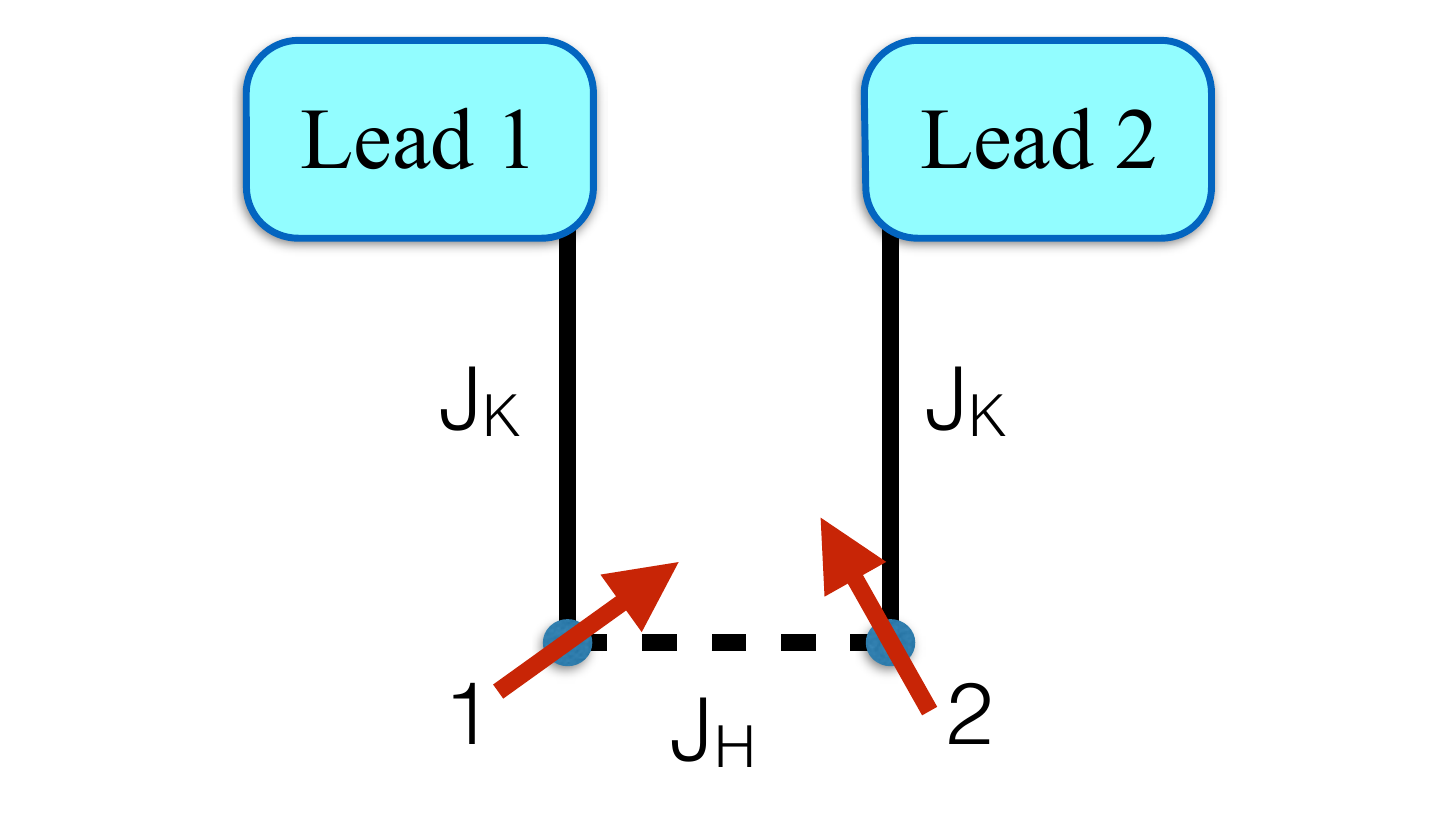}
\caption{Schematic representation of the two-impurity model.}
\end{center}
\end{figure}
The Hamiltonian is written as:
\begin{eqnarray}\label{Hc}
H=H_c + J_K\!\sum_{a,\alpha\beta} s_{a\alpha\beta}(0) S_{a\beta\alpha}+ J_H \! \sum_{\alpha\beta}  S_{1\alpha\beta}S_{2\beta\alpha},
\end{eqnarray}
where $H_c=\sum_{a\bk\sigma} \epsilon_{\bk}c_{a\bk\sigma}^\dagger c_{a\bk\sigma}$ is the conduction electron Hamiltonian, $a=\{1,2\}$ is the lead and local moment index, $\bk$ the momentum and $\sigma$ the spin index, which assume values $\sigma=\{\pm1,\pm 2,...,\pm N/2\}$ in the large-N limit. Here $s_a(0)$ is the spin density of conduction electrons at the site which is connected to the local moment spin $S_a$.  Introducing the supersymmetric-symplectic spin, the Hamiltonian can be written in the large-$N$ limit in terms of fermionic and bosonic operators as:
\begin{eqnarray}
H=H_c&-&
\frac{2J_{K}}{N}\sum_{a,\alpha \beta}{\rm Str}\left[
\left(\Psi_{a\alpha }c\dg_{a\alpha } \right)
\left(c_{a\beta}\bar \Psi_{a\beta } \right) \right]\cr
&-&\frac{J_{H}}{N}\sum_{\alpha\beta } 
{\rm Str}\left[
\left(\Psi_{1\beta }  \bar \Psi_{2\beta}  \right)
\left(\Psi_{2\alpha}
\bar  \Psi_{1\alpha } \right)
\right].\end{eqnarray}
where $\bar{\Psi}_\sigma= \Psi_\sigma^\dagger \gamma_0$, as defined in Eq.~\ref{Phi}.

We can now apply the formalism introduced in the previous section. We perform a Hubbard-Stratonovich transformation to decouple the interacting terms in the Hamiltonian by introducing fluctuating fields. Within a static mean field solution the fermionic and bosonic problems decouple and are effectively linked only by the constraint; we can now factor the partition function as:
\begin{eqnarray}
Z[q_{F}]=Z_F (q_{F}) Z_B (q_{0}-q_{F}) ,
\end{eqnarray}
where we have defined $q_{0}=Q_{0}/N$, $q_{F}=Q_{F}/N$. 
We see solutions where the Free energy is minimized with respect to
$q_{F}$. 
 
The fermionic part of the partition function reads:
\begin{eqnarray}
Z_F&=&\int \mathcal{D}\mu_F e^{-S_F}, \hspace{0.5cm} \mathcal{D}\mu_F = \mathcal{D}[c,  f]\nonumber
\end{eqnarray}
\begin{eqnarray}
S_F &&=S_c+\int_0^\beta d\tau \sum_{a,\sigma}\left[ f^\dagger_{a\sigma}(\partial_\tau+\lambda_{F}) f_{a\sigma}\right.\nonumber\\
&&\left. +\sum_{\bk}\left( f_{a\sigma}^\dagger v_{a} c_{a\bk\sigma}+ h.c.\right)\right]\nonumber\\
&&+\beta N \sum_a\frac{ |v_{a}|^2}{J_K}-2\beta N  \lambda_{F} q_F,
\end{eqnarray}
where we already dropped the terms in $p_a$ and $t_a$, since it can be shown that these do not contribute to the saddle point solution. Also, the f-operators can be redefined to eliminate the pairing term between c- and f-operators from the Hamiltonian:
\begin{eqnarray}
f_{a\sigma}\rightarrow \tilde{f}_{a\sigma}= \frac{\bar{v}_a f_{a\sigma}+d_a \tilde{\sigma} f_{a-\sigma}^\dagger}{\sqrt{|v_{a}|^2+|d_{a}|^2}}.
\end{eqnarray}

Under these considerations, the fermionic part of the solution reduces to two decoupled impurity problems. Taking $v_{a}= v$ to be site independent, integrating out the conduction electrons in each lead and transforming from imaginary time to Matsubara frequencies (see details in Appendix~\ref{AppFE}):
\begin{eqnarray}
&&S_{F}=\sum_{n a\sigma} f^\dagger_{a\sigma}(i\omega_n)(-i\omega_n+\lambda_{F}+i\Gamma_n) f_{a\sigma}(i\omega_n)\nonumber\\&&\hspace{0.8cm}+2\beta\frac{N |v|^2}{J_K}-2\beta N \lambda_{F} q_F,\\
&&\Gamma_n=\Gamma \Theta(D-|i\omega_n|)sgn (i\omega_n) ,\hspace{0.5cm}\Gamma=\pi \rho_0 |v|^2,
\end{eqnarray}
where $\rho_0$ is a constant DOS, $D$ is the bandwidth and $\Theta(x)$ is a Heaviside step function. Summing over Matsubara frequencies, in the limit $T\rightarrow 0$, the free energy has the form:
\begin{eqnarray}\label{FE0}
\frac{F_F}{N}&=&\frac{2}{\pi} Im\left[ \xi_F \ln \left(\frac{\xi_F}{eT_K e^{i\pi q_F}}\right)\right],
\end{eqnarray}
where we define
\begin{eqnarray}
\xi_F&=&\lambda_F+i\Gamma,
\end{eqnarray}
and the Kondo temperature
\begin{eqnarray}
T_K=D e^{-1/\rho_0 J_K}.
\end{eqnarray}

In the large-$N$ limit the partition function is dominated by the saddle point. Minimizing the free energy with respect to $\xi_F$ one finds 
\begin{equation}\label{LF}
\xi_F=T_Ke^{i\pi q_F},
\end{equation}
and substituting back into the fermionic free energy:
\begin{equation}
\frac{F_F}{N}=-\frac{2}{\pi} T_K \sin(\pi q_F).
\end{equation}

The bosonic part of the partition function can be concisely written as:
\begin{eqnarray}
Z_B&=&\int \mathcal{D}\mu_B e^{-S_B}, \hspace{0.5cm} \mathcal{D}\mu_B = \mathcal{D}[b, g,  \lambda_{B}]\nonumber
\end{eqnarray}
\begin{eqnarray}
S_B&=&\int_0^\beta d\tau \sum_{\sigma} \Psi_{B\sigma}^\dagger L_B \Psi_{B\sigma}\nonumber\\&&+\beta N \frac{ |g|^2}{J_H}-2\beta N\lambda_{B}( q_B+1/2),
\end{eqnarray}
where
\begin{eqnarray}
L_B=\left( \begin{array}{cc}
\partial_\tau+\lambda_B& g   \\
\bar{g} & -\partial_\tau+\lambda_B \\
\end{array} \right),
\end{eqnarray}
\begin{eqnarray}
\Psi_{B\sigma}=\left( \begin{array}{c}
b_{1\sigma} \\
\tilde{\sigma} b_{2-\sigma}^\dagger\\
\end{array} \right).
\end{eqnarray}
Here we already dropped the fluctuating field $q$ since the saddle point solution results in $q=0$. Integrating out the bosons and summing over Matsubara frequencies (see Appendix~\ref{AppBE}), in the zero temperature limit, the free energy is given by:
\begin{eqnarray}
\frac{F_B}{N}&=&\sqrt{\lambda_B^2-|g|^2} + \frac{ |g|^2}{J_H}-2\lambda_{B} (q_B+1/2).
\end{eqnarray}

Minimizing the free energy with respect to $g$ and $\lambda_B$ one finds:
\begin{equation}\label{LB}
\lambda_B=J_H(q_B+1/2), \hspace{1cm}
|g|^2=J_H^2q_B(q_B+1),
\end{equation}
so the the bosonic free energy can be written as:
\begin{equation}
\frac{F_B}{N}=-J_H(q_B+1/2)^2,
\end{equation}
up to a constant term.

%%%%%%%%%%%%%%%%%%%%%
\subsection{Analysis of the free energy}

\begin{figure}[t]
\begin{center}
\includegraphics[width=0.95\linewidth, keepaspectratio]{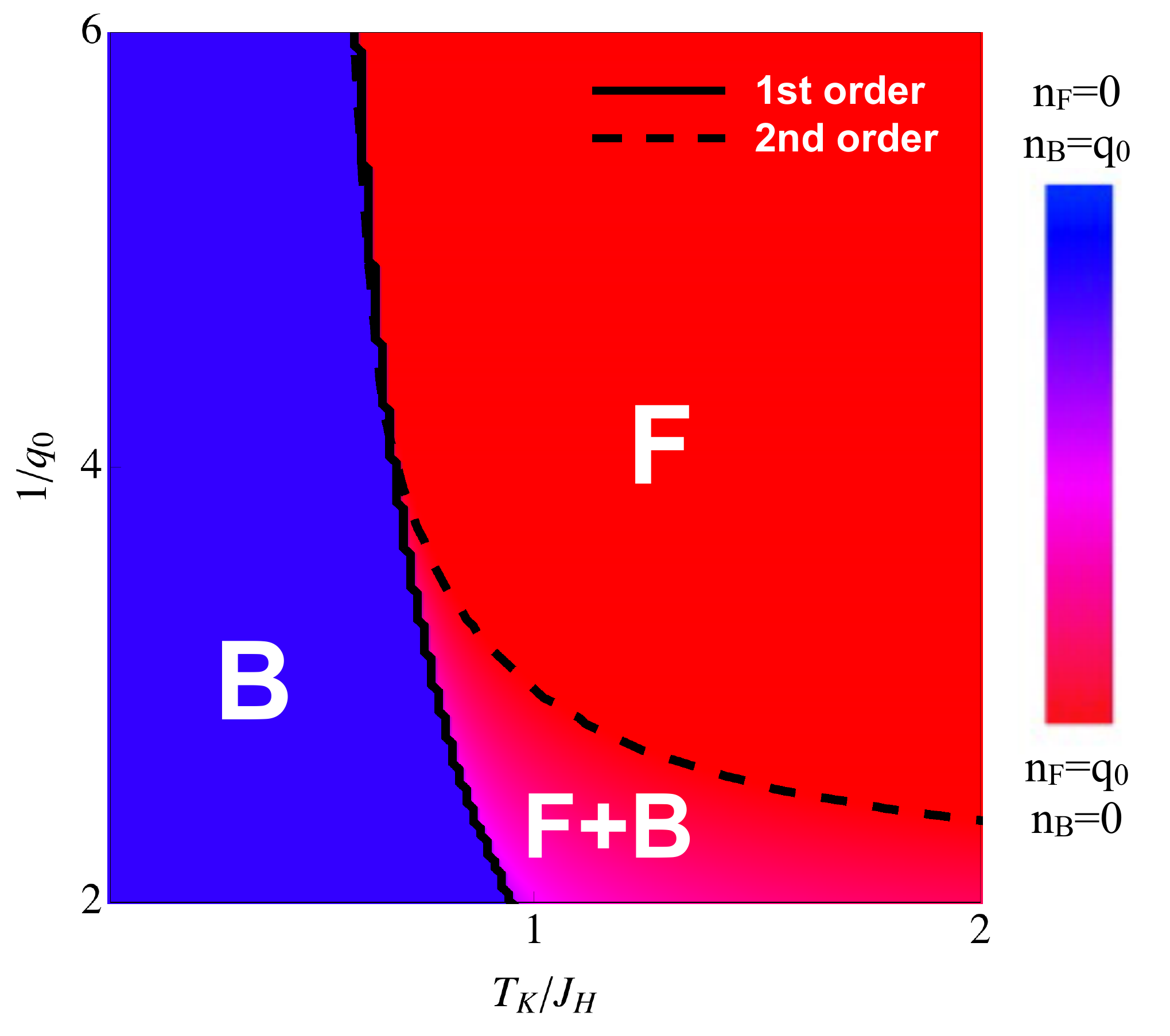}
\caption{Representation diagram for the two impurity model indicating the most favorable representations as a function of $A=\frac{T_K}{J_H}$ and $1/q_0$. A purely fermionic representation phase (red) is labeled by $F$, a pure bosonic representation phase (blue) is labeled as $B$ and the mixed representation phase (intermediate colors) is labeled by $F+B$.}
\label{PD}
\end{center}
\end{figure}

%The total free energy
Making explicit use of the constraint condition, $q_F+q_B=q_0$, the total free energy can be written as:
\begin{eqnarray}
\frac{F}{J_H N}=-\frac{2}{\pi} A \sin(\pi q_F)- (q_0-q_F+1/2)^2,
\end{eqnarray}
where $A= \frac{T_K}{J_H}$ and the free energy is given in units of
$J_H$. For each value of $A$ and $q_0$ the representation was
determined by the minimization of the free energy with respect to
$q_F$ and the result is plotted in the \emph{representation diagram} of Fig.~\ref{PD}. In case the free energy is minimized for $q_F=q_0$
the phase is purely fermionic, meaning that a completely antisymmetric
representation is favored. Analogously, for $q_F=0$ (or $q_B=q_0$) the
phase is purely bosonic, and a symmetric representation is more
appropriate. Solutions with $0<q_F<q_0$ are solutions in which both
bosons and fermions coexist, which we call a \emph{mixed phase} and
label as $(F+B)$ in Fig.~\ref{PD}.  Note that for a fixed value of
$q_0$, as the ratio $T_K/ J_H$ is increased the spins tend to develop
fermionic character. Also, for a fixed value of $T_K/J_H$, increasing
$1/q_0$ (or reducing $q_0$, which is equivalent to decreasing the
magnitude of the spin) the spin representation also tends towards a
fermionic representation.

The dashed line between $F$ and $F+B$ regions indicates a second order phase transition  and can be determined from the condition:
\begin{eqnarray}
%\frac{F[q_F=q_0]}{N}=\frac{F[q_F=q_0-\delta]}{N}\Bigg|_{\delta\rightarrow
%0}
\left. \frac{\partial F}{\partial q_{F}}\right\vert_{q_{F}=q_0} = 0
,
\end{eqnarray}
which leads to:
\begin{eqnarray}\label{2nd}
A=\frac{1}{2 \cos(\pi q_0)}.
\end{eqnarray}

The continuous line represents a first order phase transition. The line between the purely fermionic and purely bosonic representations is determined by:
\begin{eqnarray}
\frac{F[q_F=q_0]}{N}=\frac{F[q_F=0]}{N},
\end{eqnarray}
which gives the condition:
\begin{eqnarray}
A=\frac{\pi q_0(q_0+1)}{2 \sin(\pi q_0)}.
\end{eqnarray}
The first order line between the phases $B$ and $F+B$ cannot be computed analytically and was determined numerically.

Throughout the mixed phase we have $\lambda_F=\lambda_B$, so both
fluids have the same chemical potential. This is related to the
presence of the Type II minima of the free energy (see discussion in
the introduction), with a saddle point that allows the coexistence of
bosons and fermions and the interchange of one into another at no
energy cost. In particular, at the second order ``phase transition''
line discussed above, when $q_F=q_0-\delta$ and $q_B=\delta$ with
$\delta\rightarrow 0$, one can check explicitly from Eq.~\ref{LB} and
\ref{LF} that the condition $\lambda_F=\lambda_B$ gives the same
condition that defines the second order line in Eq.~\ref{2nd}.

%We are able to find that the character of the spin changes as one samples the phase diagram as a function of the ratio of $T_K/\pi J_H$, being purely bosonic for large $J_H$, and fermionic or in a mixed representation for large $T_K$. The mixed representation of the spin can provide a description for coexistence of magnetism and superconductivity or heavy particles very naturally. In addition, we find a second order phase transition at which a fermionic mode can potentially go soft in lattice models, which can lead to non Fermi liquid behavior.

%It can be found by analyzing the free energy $F(q_F,Q_B)$ under the condition $F(q_0,0)=F(q_0-\delta,\delta)|_{\delta\rightarrow 0}$, which defines the transition at $A=2\pi \cos(\pi q_0)$. The first order transition line (between $B$ and $F$) is defined by the condition $F(q_0,0)=F(0,q_0)$, which gives the transition at $A=\frac{2\sin(\pi q_0)}{q_0(q_0+1)}$.

%***Add discussion about the VJ fixed point. Maybe add the point to the phase diagram?

%%%%%%%%%%%%%%%%%%%%%
\subsection{Fluctuations of the local fermionic fields}

We now analyze the effects of fluctuations of the local fermionic fields. In the previous section we introduced the time dependent fields $\phi_{a}$ and $\xi_a$, which allow us to decouple the terms $(b_{a\alpha} c_{a\bk\alpha}^\dagger)(c_{a\bk\beta} b_{a\beta}^\dagger)$ and $(\tilde{\alpha}b_{a-\alpha}^\dagger c_{a\bk\alpha}^\dagger)(\tilde{\beta} c_{a\bk\beta} b_{a-\beta})$ in the action, respectively. These fields do not acquire an expectation value, but fluctuate around zero. The partition function can now be written in terms of the saddle point solution determined in the former subsection times $Z_{\phi}$ and $Z_{\xi}$, the new  contributions to the partition function due to the presence of the fluctuating fields $\delta \phi$ and $\delta \xi$, that we take to be site independent. Focusing first in the $\delta\phi$ field:
\begin{eqnarray}
Z_{\phi}\!&=&\!\int \! D\phi e^{-\int_{0}^\beta d\tau\left[\sum_{a\bk\alpha}( b_{a\alpha}^\dagger c_{a\bk\alpha}\delta\phi+h.c.) +\frac{2N|\delta\phi|^2}{J_K}\right]}.
\end{eqnarray}

Expanding to second order in $\delta\phi$ we can identify the propagator for the fluctuating field $\delta\phi$:
\begin{eqnarray}
[D_\phi(i\omega_r)]^{-1}= 2N\left[ \chi_{cb}(i\omega_r) -\frac{1}{J_K}\right],
\end{eqnarray}
with
\begin{eqnarray}
\chi_{cb}(i\omega_r)&=&\!\!\!\!\!\vcenter{\includegraphics[width=5cm]{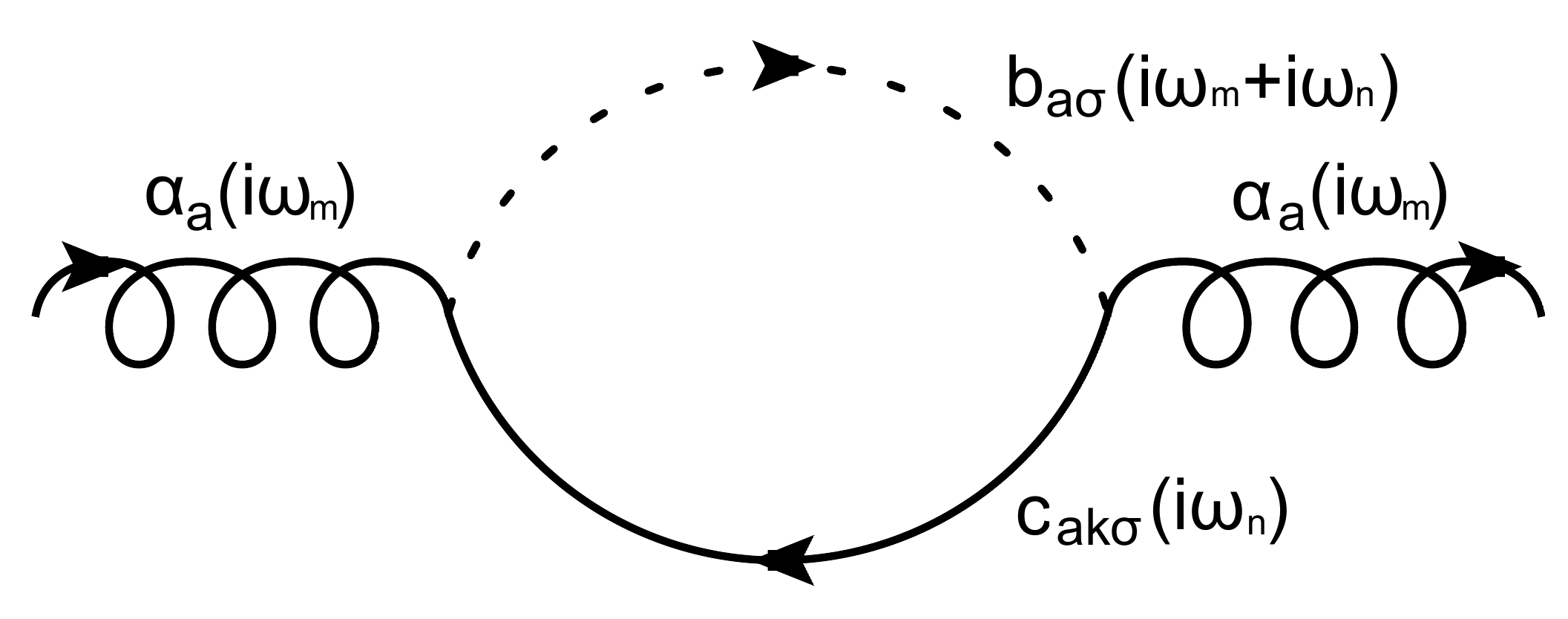}}\nonumber\\
&=& -\frac{1}{\beta}\sum_{\bk \bk' m}G_{b} (i\omega_{r}+i\omega_m) G_{\bk\bk'}(i\omega_m),\nonumber
\end{eqnarray}
where $G_b(i\nu_n)$ is the bosonic propagator and $G_{\bk\bk'}(i\omega_n)$ is the full c-electron propagator. 

We now evaluate $\chi_{cb}(\omega)$ (details of the calculation can be found in Appendix~\ref{AppFlu}). One interesting region for the analysis of $D_\phi(i\omega_r)$ is the second order transition line, where the energy levels of the bosons and fermions are equal. In the infinite bandwidth limit we find:
\begin{eqnarray}
\chi_{cb}(\omega -i\delta)-\frac{1}{J_K}&=& \rho_0 \omega Re\left[ \frac{ \log(\lambda+i\Delta)}{\omega+i\Delta}\right]\nonumber\\&-&\frac{\rho_0  \omega ^2}{\Delta ^2+\omega ^2}\log (\lambda -\omega + i\delta).
\end{eqnarray}
Note that at zero frequency, $\chi_{cb}(0)-\frac{1}{J_K}=0$, so
$[D_\phi(0)]^{-1}= 0$, and the propagator for the fermionic
hybridization field $\phi$ diverges at zero frequency at the second order phase
transition, indicating the presence of a fermionic zero mode. Also,
there is a gap of magnitude equal to $\lambda=\lambda_F=\lambda_B$
with a continuum that goes up to the bandwidth. This gap is always
present in the 2-impurity model since $\lambda=\xi_B=J_H/2$ is aways
finite at the transition. For a Kondo-Heisenberg model in the lattice,
the bosonic level will acquire a dispersion and when magnetic order
sets in it will be gapless at some points in the Brillouin zone. In
that case the spectrum for the fermionic hybridization field is
expected to have a continuum of excitations, which can potentially
lead to non Fermi liquid behavior.

\begin{figure}[H]
\begin{center}
\includegraphics[width=0.8\linewidth, keepaspectratio]{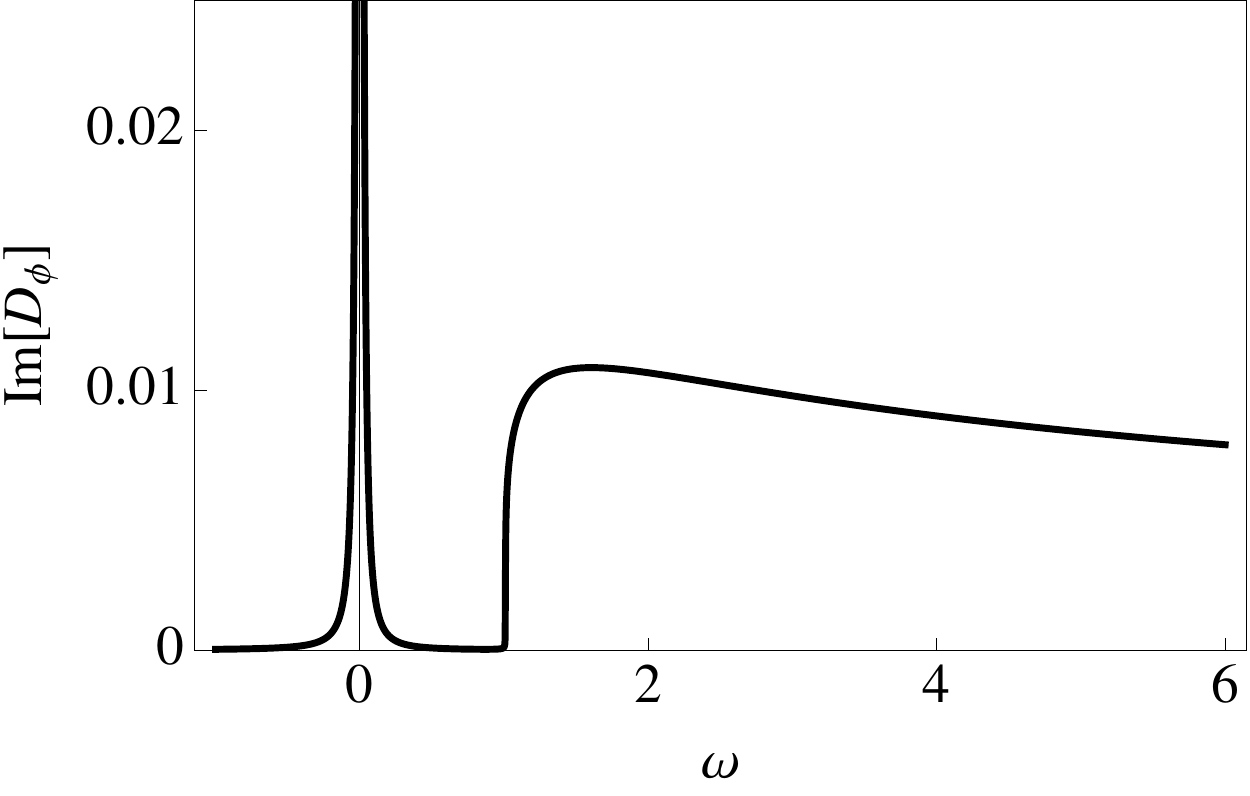}
\caption{Plot of the imaginary time of $D(\omega)$, the $\delta\phi$ propagator. The parameters used in this plot were a solution of the mean field theory at a specific point of the second order phase transition: $A=1.57$ and $1/q_0=2.5$, which gives $\lambda_F=\lambda_B=\lambda=1.01$ and $\Delta=2.97$ (in units of $J_H$). }
\label{ImD}
\end{center}
\end{figure}

For the second fermionic mode $\delta \xi$ a similar calculation follows, where now:
\begin{eqnarray}
[D_\xi(i\omega_r)]^{-1}= 2N\left[ \bar{\chi}_{cb}(i\omega_r) -\frac{1}{J_K}\right],
\end{eqnarray}
with
\begin{eqnarray}
\bar{\chi}_{cb}(i\omega_r)&=&\!\!\!\!\!\vcenter{\includegraphics[width=5cm]{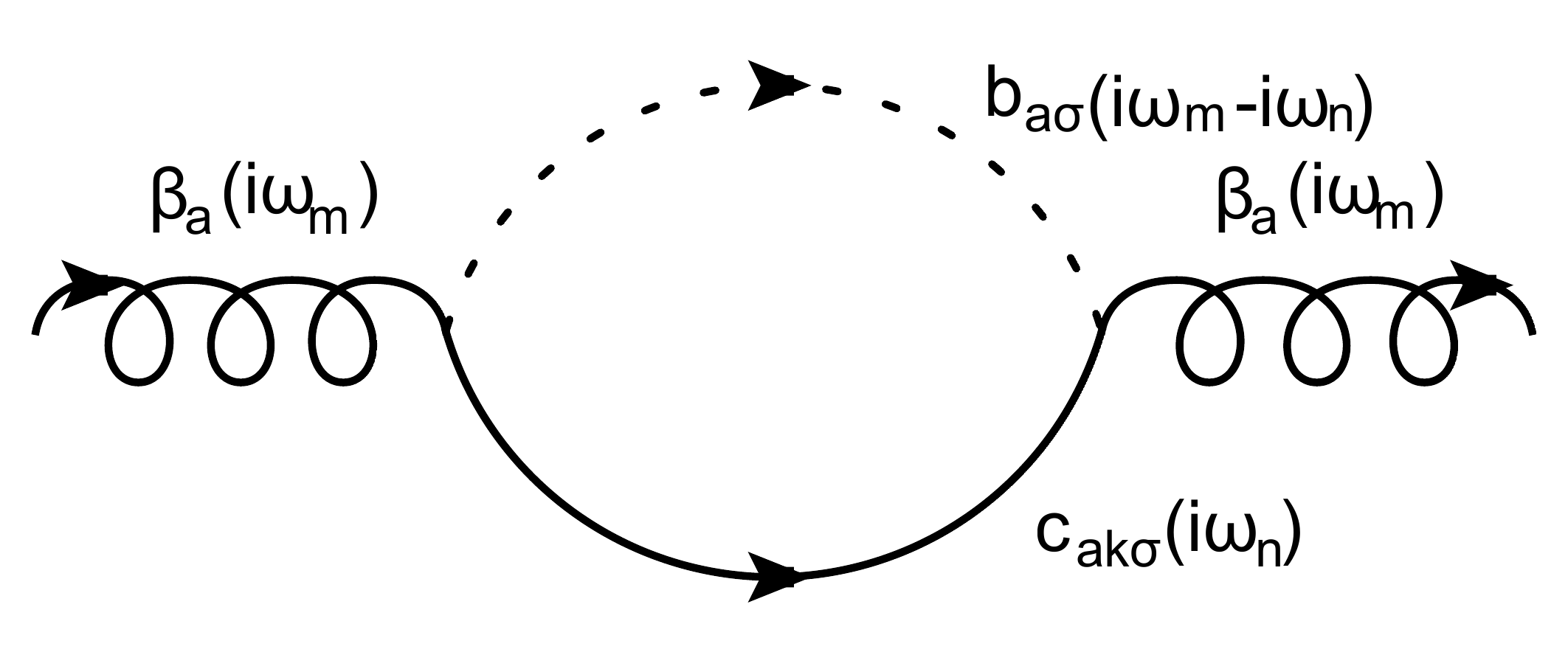}}\nonumber\\
&=& \frac{1}{\beta}\sum_{\bk \bk' m}G_{b} (i\omega_{r}-i\omega_m) G_{\bk\bk'}(i\omega_m),\nonumber
\end{eqnarray}
and we can write, at the second order transition line:
\begin{eqnarray}
\bar{\chi}_{cb}(\omega -i\delta)&-&\frac{1}{J_K}= \rho_0 (\omega-2\lambda) Re\left[ \frac{ \log(-\lambda+i\Delta)}{\omega-2\lambda+i\Delta}\right]\nonumber\\&-&\frac{\rho_0  (\omega-2\lambda)^2}{\Delta ^2+(\omega-2\lambda)^2}\log (\lambda -\omega + i\delta).
\end{eqnarray}

Here we note that there is no zero mode for this fermionic field at the transition, and a continuum starts at a finite $\lambda=\xi_B=J_H/2$. This is related to the choice we made to implement the constraint, which is not invariant under all transformations that leave the spin invariant. In particular, it is not invariant under transformations of the form $g_B$ (see Appendix~\ref{AppSS}), generated by the operators $\eta^\dagger$ and $\eta$.

%Interestingly enough, in the limit of $\lambda\rightarrow 0$, when we recover particle-hole symmetry the propagators for the $\phi$ and $\xi$ fields actually coincide and there are two zero modes. That can be checked by assuming particle-hole symmetry: $G_{\bk\bk'}(-i\omega)=-G_{\bk\bk'}(i\omega)$, in which case:\begin{eqnarray} \bar{\chi}_{cb}(i\omega_m)&=&\frac{1}{\beta}\sum_{n\bk\bk'}G_{\bk\bk'}(i\omega_n) G_{b}(i\omega_m-i\omega_n),\\&=&\frac{1}{\beta}\sum_{n\bk\bk'}G_{\bk\bk'}(-i\omega_n) G_{b}(i\omega_m+i\omega_n),\nonumber\\&=&-\frac{1}{\beta}\sum_{n\bk\bk'}G_{\bk\bk'}(i\omega_n) G_{b}(i\omega_m+i\omega_n),\nonumber\\&=&\chi_{cb}(i\omega_m).\nonumber\end{eqnarray}

%%%%%%%%%%%%%%%%%%%%%%%%%%%%%%%%%%%%%%%%%%%%%%

\section{Frustration in the three impurity model}\label{Sec3Imp}

\begin{figure}[b]
\begin{center}
\includegraphics[width=0.8\linewidth, keepaspectratio]{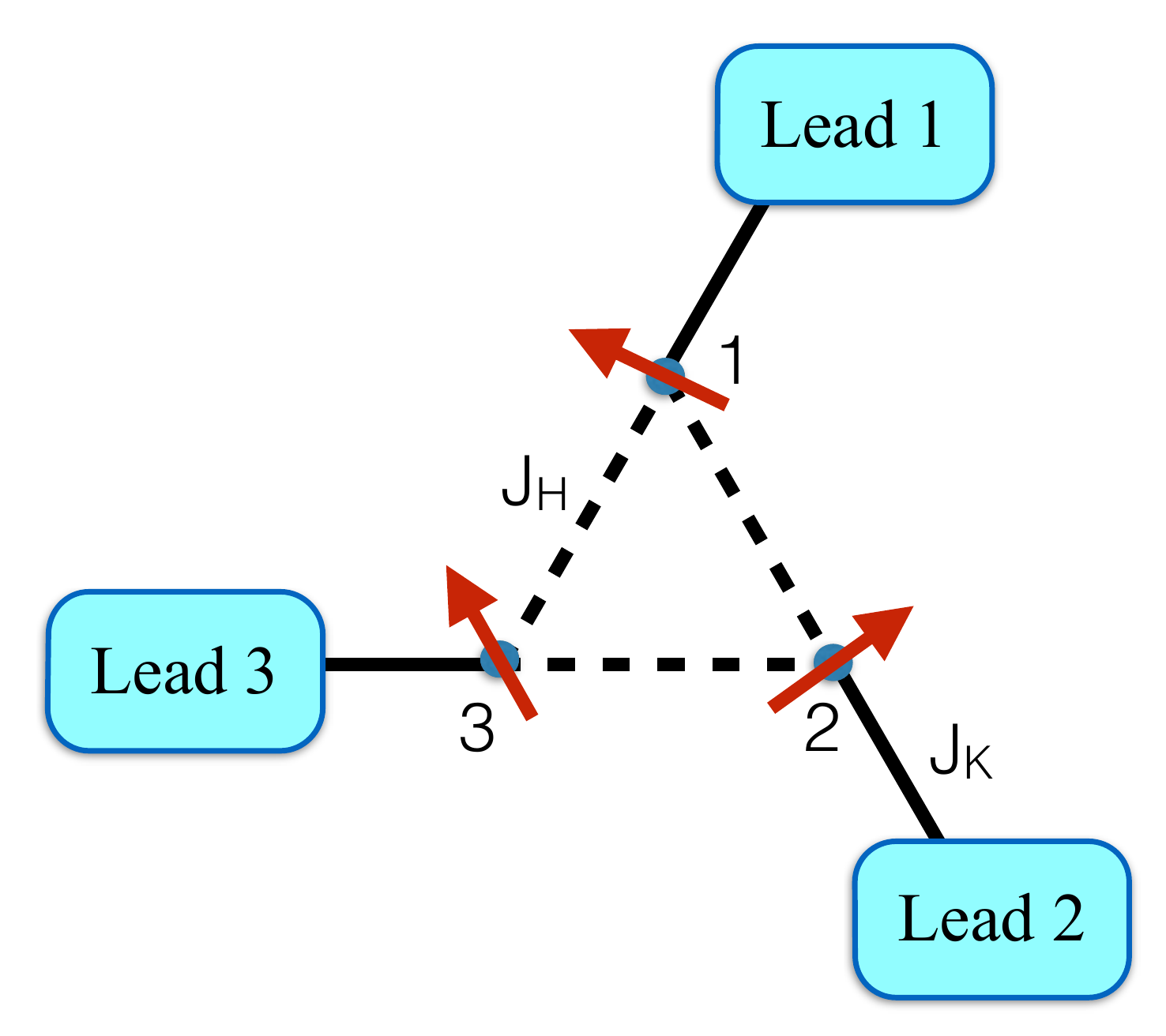}
\caption{Schematic representation of the frustrated three-impurity model.}\label{Tri}
\end{center}
\end{figure}

%The model %Writing in terms of B and F
As a second application of the supersymmetric-symplectic spin, we study a minimal model
that brings in the issue of geometric frustration into play. The model
consists of three local moments interacting among themselves by an
antiferromagnetic Heisenberg coupling $J_H$ and interacting with its
respective bath of conduction electrons by a Kondo coupling $J_K$, as
depicted in Fig.~\ref{Tri}. We are motivated to look at this problem
by experiments in CePdAl in which the equivalent Ce sites
spontaneously develop a state in which one third are paramagnetic and the
other two thirds are magnetically ordered
\cite{Fri2}, exploring the ability of the symplectic representation of the spin to describe frustrated systems.

The Hamiltonian is written as:
\begin{eqnarray}
H=H_c + J_K\!\sum_{a,\alpha\beta} s_{a\alpha\beta}(0)S_{a\beta\alpha}+ J_H \!\! \sum_{a,\alpha\beta}\!\!S_{a\alpha\beta}S_{a+1\beta\alpha},\nonumber\\ 
\end{eqnarray}
where now $a=\{1,2,3\}$ is the lead and local moment index with periodic boundary conditions. As in the previous section, $H_c$ is the conduction electron Hamiltonian, $s_a(0)$ is the spin density of conduction electrons at the site that is connected to the local moment spin $S_a$. 

Introducing the supersymmetric-symplectic spin from Eq.~\ref{S} into the Hamiltonian, this can be written in the large-$N$ limit as:
\begin{eqnarray}
H&=&H_c-
\frac{2J_{K}}{N}\sum_{a,\alpha \beta}{\rm Str}\left[
\left(\Psi_{a\alpha }c\dg_{a\alpha } \right)
\left(c_{a\beta}\bar \Psi_{a\beta } \right) \right]\cr
&-&\frac{J_{H}}{N}\sum_{a,\alpha \beta } 
{\rm Str}\left[
\left(\Psi_{a\beta }  \bar \Psi_{a+1\beta}  \right)
\left(\Psi_{a+1\alpha}
\bar  \Psi_{a\alpha } \right)
\right].
\end{eqnarray}
with $\Psi_{a\sigma}$ defined in Eq.~\ref{Phi}.

Proceeding as in the previous section, within a path integral
formalism we introduce fluctuating fields in oder to decouple the
quartic terms in the action and impose the constraint by the
introduction of a delta function in the integral form. Within a
static saddle point solution the problem decouples in to a bosonic and
a fermionic part, effectively linked by the constraint. In this
section we are going to leave the representation of the spin and the
mean field parameters to be determined independently in each
site. Omitting the details (similar to the previous section), the
partition function in the large $N$ limit can be written as 
\begin{eqnarray}
Z=Z_F Z_B,
 \end{eqnarray}
with the understanding that the partition function is to be stationary
with respect to the $q_{Fj}$ at the three sites. 

%Note now that the saddle point solution admits $q\neq 0$. form a singlet between leads $a$ and $a+1$, taking care of the antiferromagnetic correlations. The term in $B^\dagger_a$ generates hopping of the bosons between leads, which tends to destroy the AF correlations, naturally implementing the frustrated character of the triangle geometry into the problem.
The fermionic part of the partition function can be written as:
\begin{eqnarray}Z_F&=&\int \mathcal{D}\mu_F e^{-S_F}, \hspace{0.5cm} \mathcal{D}\mu_F = \mathcal{D}[c, f]\nonumber\end{eqnarray}
\begin{eqnarray}S_F &&=S_c+\int_0^\beta d\tau \sum_{a,\sigma} \left[\fb_{a\sigma}(\partial_\tau+\lambda_{Fa}) f_{a\sigma}\right.\nonumber\\&&\left. +\sum_{\bk}\left(f^\dagger_{a\sigma}v_{a}c_{a\bk\sigma}+ h.c.\right)\right]\nonumber\\&&+\beta N \sum_{a}\frac{ |v_{a}|^2}{J_K}-\beta N \sum_a \lambda_{Fa} q_{Fa},\end{eqnarray}
where as in the previous section, we assume that the only fluctuating field that acquires a finite value at the saddle point solution is $v_{a}$. In this case the fermionic part of the solution reduces to three decoupled impurity problems, with the same solution as the previous section, now for 3 leads:
\begin{equation}
\frac{F_F}{N}=-\frac{T_K}{\pi}  \sum_a\sin(\pi q_{Fa}).
\end{equation}

The bosonic part of the partition function reads:
\begin{eqnarray}
Z_B&=&\int \mathcal{D}[b]e^{-S_B}, 
\end{eqnarray}
where 
\begin{eqnarray}
S_B&=&\int_0^\beta d\tau \sum_{\sigma} \Psi_{B\sigma}^\dagger\frac{L_B }{2}\Psi_{B\sigma} \nonumber\\&-&\frac{\beta N}{2J_H}\sum_aTr[\Delta^\dagger_{Ba}\gamma_0^B \Delta_{Ba} \gamma_0^B]\nonumber\\&-&\beta N\sum_a\lambda_{Ba}( q_{Ba}+1/2),
\end{eqnarray}
where we denote $q_{Ba}=q_{0}-q_{Fa}$ and 
\begin{eqnarray}
L_B=\left( \begin{array}{ccc}
m_{B1} &\Delta_{B1}& \Delta_{B3}^\dagger\\
\Delta_{B1}^\dagger & m_{B2}&  \Delta_{B2}\\
\Delta_{B3} & \Delta_{B2}^\dagger &m_{B3}
\end{array} \right),
\end{eqnarray}
with
\begin{eqnarray}
m_{Ba}=\left( \begin{array}{cc}
\partial_\tau +\lambda_{Ba}&0 \\
0 & -\partial_\tau +\lambda_{Ba} \end{array} \right),
 \end{eqnarray}
 \begin{eqnarray}
\Delta_{Ba}=\left( \begin{array}{cc}
q_a & -g_a \\
 \bar{g}_a & \qb_a \end{array} \right),
 \end{eqnarray}
and
\begin{eqnarray}
\Psi_{B\sigma}=\left( \begin{array}{c}
b_{1\sigma} \\
\tilde{\sigma} b_{1-\sigma}^\dagger\\
b_{2\sigma} \\
\tilde{\sigma} b_{2-\sigma}^\dagger\\
b_{3\sigma} \\
\tilde{\sigma} b_{3-\sigma}^\dagger\\
\end{array} \right).
\end{eqnarray}
Note that the trace term in the action now appears with a minus sign since it is related to the bosonic part of the super-trace introduced in Eq.~\ref{Str}. Here we define $\gamma_0^B=\sigma_3$ as the bosonic part of the original matrix $\gamma_0$.

The solution of the bosonic part of the partition function is more involved if we allow the representations of the spin to be different in each lead. As a first solution we consider the same representation on every site, and look for an \emph{homogeneous solution}, taking $\lambda_{Ba}\rightarrow\lambda_B$, $q_a\rightarrow q$ and $g_a\rightarrow g$. Integrating out the bosons and summing over Matsubara frequencies, in the zero temperature limit, the free energy can be written as:
\begin{eqnarray}
\frac{F_B}{N}&=&(\lambda_B+2q)+2\sqrt{(\lambda_B-q)^2-3g^2}\nonumber\\ &&+\frac{3(g^2-q^2)}{J_H}-3\lambda_B(q_B+1/2).
\end{eqnarray}

Minimizing the free energy with respect to $q$, $g$ and $\lambda_B$ one finds the saddle point free energy:
\begin{equation}
\frac{F_B}{N}=-\frac{3J_H}{2}(q_B+1/2)^2,
\end{equation}
up to a constant term.  The total free energy for the homogeneous
solution can be written, already making explicit use of the constraint
condition $q_F+q_B=q_0$, as:
\begin{eqnarray}
\frac{F_{Hom}}{J_H N}=- \frac{3}{\pi} A \sin(\pi q_F)-\frac{3}{2}(q_0-q_F+1/2)^2,
\end{eqnarray}
where again $A= \frac{T_K}{J_H}$ and the free energy is given in units of $J_H$. Note that this is functionally the same as the 2-impurity model up to an overall factor of $3/2$, and as a consequence the representation diagram determining the most favorable representation for the spin within an homogeneous solution will be identical to the 2-impurity case. 

\begin{figure}[t]
\begin{center}
\includegraphics[width=0.95\linewidth, keepaspectratio]{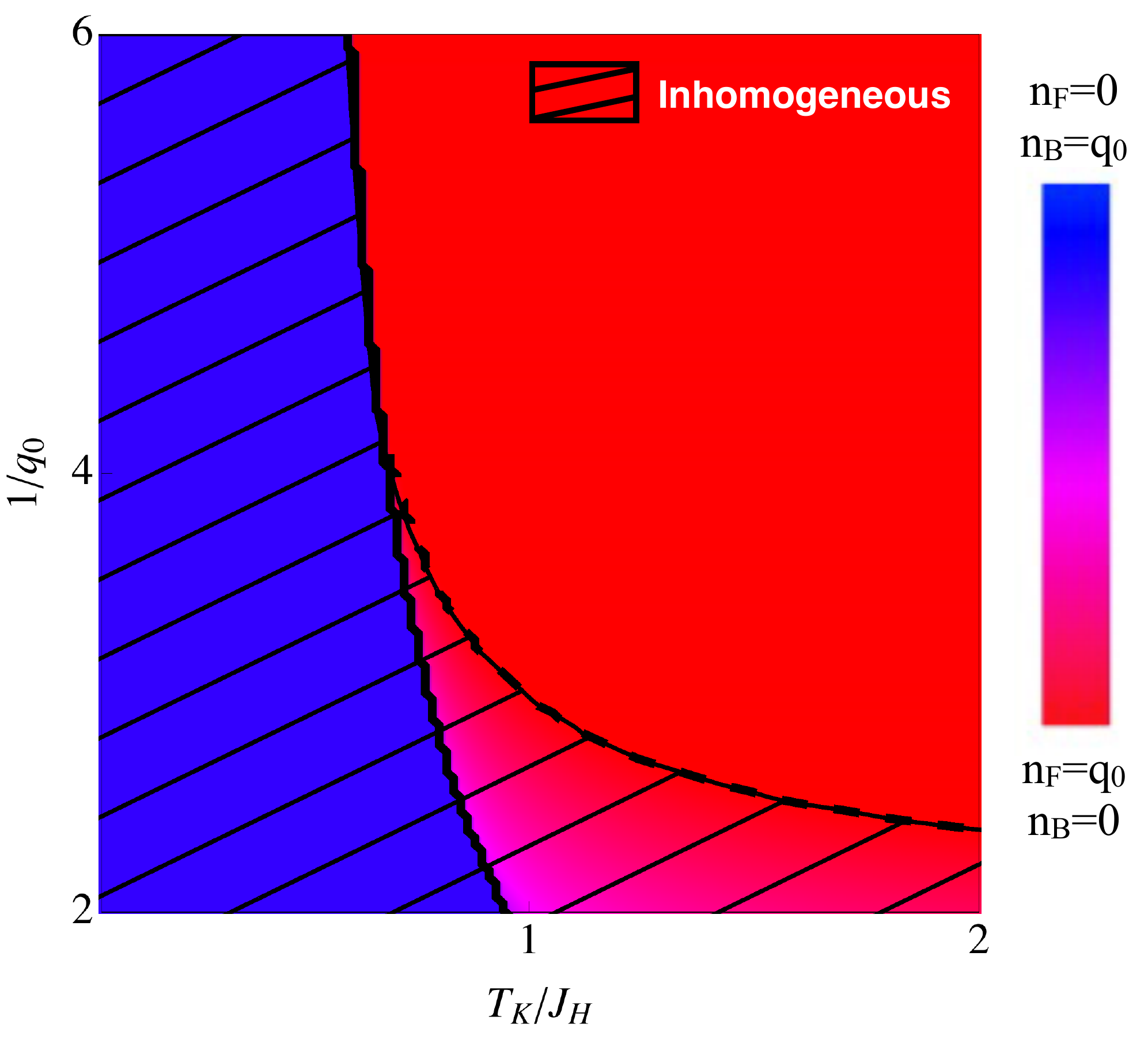}
\caption{Representation diagram for the three impurity model indicating the most favorable representations as a function of $A=\frac{T_K}{J_H}$ and $1/q_0$. The color code is the same as in Fig.~\ref{PD}. The hashed area represents the region of the diagram in which an inhomogeneous solution is energetically favorable.}
\label{PD2}
\end{center}
\end{figure}

Now we move on to investigate solutions which spontaneously develop
different representations in each site, which we refer to 
as \emph{inhomogeneous representations}. Due to frustration, we expect
that it is energetically favorable for one of the spins to be in a
fermionic representation, essentially disconnected from the other two
spins with a bosonic or mixed representation, forming an
antiferromagnetic bond. We assume one of the spins to always have a
fermionic representation and let the representation of the two other
spins to be selected as the one that minimizes the total energy.

Now the problem reduces to a single impurity problem with a purely fermionic representation plus the two-impurity problem solved in the previous section. The free energy for the inhomogeneous solution reads:
\begin{eqnarray}
\frac{F_{Inh}}{J_H N}&=&-\frac{1}{\pi}A\sin(\pi q_0)-\frac{2}{\pi} A \sin(\pi q_F)\\&-&(q_0-q_F+1/2)^2,\nonumber
\end{eqnarray}
in units of $J_H$, where $A= \frac{T_K}{J_H}$. Again, the representation diagram
 will be the same as before, but now we compare the free energies of
the homogeneous and inhomogeneous solutions for the 3-impurity
problem. The hashed area in Fig.~\ref{PD2} is the region of the
diagram in which the inhomogeneous solution is more
favorable. This result provides a model for the situation which
appears to occur 
in CePdAl \cite{Fri2}: one third of the
local moments in the frustrated Kagome lattice (formed by an assembly
of corner-shared triangles) relieve the frustration by
assuming an antisymmetric character and forming a Kondo singlet with a
conduction electron, while the other two thirds of the local moments
assume a bosonic character, developing magnetic order,
allowing a partially ordered phase to be formed, as depicted in
Fig.~\ref{Inhom}.

%The study of a three-impurity model also brings insight into the understanding of frustrated systems.   By the use of symplectic spins the model includes antiferromagnetic bonds and ferromagnetic correlations between them, properly accounting for frustration. We find that inhomogeneous representations of the spin are actually favorable when frustration increases, here related to decreasing the size of the spin and increasing quantum fluctuations. This result motivates further studies in the Kagome lattice in order to understand the partially disordered phase verified experimentally in CePdAl.

%%%%%%%%%%%%%%%%%%%%%%%%%%%%%%%%%%%%%%%%%%%%%%

\section{Conclusion and discussion}\label{SecCon}

%Which we find
In this work we introduced a new supersymmetric-spin representation
for large-$N$ treatments, based on a symplectic generalization of the spin operator. We have analyzed the properties of the
supersymmetric-symplectic spin and its symmetries, identifying 
the supergroup $SU(2|1)$ of transformations under which the spin is invariant.

We have proposed a new framework in the large-$N$ limit, which allows
the problem to sample different representations, selecting the one 
which lowers the energy in a given point in parameter space. This
opens up the possibility of describing the phase diagram of heavy
fermions within a single approach that can capture the
evolving character of the spin, at the same time that it offers a potential 
framework for the phenomenological two-fluid picture for heavy
fermions.

Applying this approach to two toy models, the two-impurity model and
the frustrated three-impurity model, we have shown two new classes of
mean-field theory that may be of interest in developing 
a unified description of heavy fermion systems. In particular, 
we find a \emph{mixed
phase} solution in which bosons and fermion coexist at each site, which points
the way to a description 
of the coexistence of magnetism
and Kondo effect. Also, we find stable \emph{inhomogenous
solutions}, which may provide a basis for describing 
the partially ordered state in CePdAl \cite{CePdAl1,CePdAl2, Fri2}. 

%Discussion and open questions
There are many open questions raised by our work. 
Firstly, within the mixed phase, we find evidence for 
a new kind of zero mode, a {\sl Goldstino}, that results from the 
partial breaking of supersymmetry. This can be understood as a
consequence of the fact that the Hamiltonian is invariant under
super-rotations while at the same time, the mean-field solution
breaks this rotational symmetry. 
Given a state $|\Psi\rangle$ with a fermion assigned to the
corner box, for example, one can rotate this state as follows:
\begin{eqnarray}
|\Psi\rangle \rightarrow \theta\dg _{j} \vert \Psi \rangle = 
f\dg _{j\sigma } b_{j\sigma }
|\Psi\rangle,
\end{eqnarray}
and find a new state which has same energy.  These kinds of zero modes
are only present when the chemical potential of the bosons and
fermions are equal, i.e in type II solutions. We note that 
in the presence
of a Kondo effect, the f-electrons are charged, whereas b-bosons 
are neutral, so the fermionic Goldstino excitation is charged
and may represent a {\sl zero energy valence fluctuation}. 
Further work is needed to establish whether such  zero modes are a truly
physical excitation and whether they participate 
in anomalous inelastic scattering and the development of non-Fermi liquid behavior.

Secondly, we would like to discuss the future application of this approach to
the exploration of the phase diagram of the Kondo lattice. 
Extrapolating the results found in this work to the
lattice we expect to find that for small ratios of $T_K/J_H$ a bosonic
representation is more favorable in which  magnetic order
will emerge once the bosons condense. 
Heavy
Fermi liquid behavior is expected to develop when there are fermions
in the representation, which can hybridize with the conduction
electrons. If a mixed phase is stable in the lattice, 
heavy fermion behavior can coexist with magnetic order. In
principle, this approach allows us to explore the different kinds of
phase transitions as seen in YbRh$_2$Si$_2$\cite{Pas,Geg2,Fri1} within a single approach:
given that the fermions now can form a Fermi surface, the magnetic
phase can emerge both from local moments within the bosonic
representation or as an instability of the large Fermi surface. 
Another aspect of interest
is the possibility of the description of
superconductivity; this phase is likely to develop in the mixed phase
and its vicinity, as a consequence of a valence-bond kind of magnetism
emerging from the fermionic antiferromagnetic bonds. This is another
interesting direction for future work.

Acknowledgments. The authors thank Catherine P\' epin, Onur Erten and
Tzen Ong for fruitful discussions and the Kavli Institute for
Theoretical Physics for hosting the authors during the Fall 2014 where
part of this work was completed.   This research was supported in part
by the National Science Foundation under Grant No. NSF PHY11-25915 and
NSF DMR-1309929 and by a Simons Foundation fellowship (PC).

\appendix
%%%%%%%%%%%%%%%%%%%%%%%%%%%%%%%%%%%%%%%%%%%%%%

\section{Details on the properties of supersymmetric symplectic spin}\label{AppSS}

In this appendix we discuss a few properties of the supersymmetric symplectic spin in more detail. 

\subsection{Number of spin components}
\label{AppA1}
When writing the components of the supersymmetric-symplectic spin as:
\begin{eqnarray}
S_{\alpha\beta}= f_{\alpha}^\dagger f_{\beta} - \tilde{\alpha}\tilde{\beta}' f_{-\beta}^\dagger f_{-\alpha} +b_{\alpha}^\dagger b_{\beta}  - \tilde{\alpha}\tilde{\beta}' b_{-\beta}^\dagger b_{-\alpha},
\end{eqnarray}
note that we have in fact only $N(N+1)/2$ components (equal to the number of generators of the enlarged symmetry group of the spin, here $SP(N)$). This can be checked by noticing that $S_{\alpha\beta}=-\tilde{\alpha}\tilde{\beta}S_{-\beta-\alpha}$, so the generators are not all independent. For $\alpha,\beta>0$, we have $S_{\alpha\beta}=-S_{-\beta-\alpha}$, so the generators with both indexes negative are linearly dependent on the generators with both indexes positive. For $\alpha>0, \beta<0, \alpha\neq -\beta$ we have $S_{\alpha\beta}=S_{-\beta-\alpha}$, but note that for the case of $\alpha=\beta$ we have the condition $S_{\alpha,-\alpha}=S_{\alpha,-\alpha}$, which does not give a relation between the generators in the off diagonal. One can check that indeed $S_{\alpha,-\alpha}$ is not linearly dependent on $S_{-\alpha,\alpha}$. Within these considerations the number of independent generators is $N(N+1)/2$, as expected.

\subsection{Symmetry transformations of the supersymmetric symplectic
spin}
\label{AppA2}
In the main text, a more concise form of the supersymmetric-symplectic spin is introduced in order to make the local symmetry manifestly clear:
\begin{eqnarray}
S_{\alpha\beta}=\bar{\Psi}^\dagger_{\alpha}  \Psi_{\beta}= \Psi^\dagger_{\alpha} \gamma_0 \Psi_{\beta},
\end{eqnarray}
where, 
\begin{eqnarray}
\Psi_{\alpha}=\left( \begin{array}{c}
f_\alpha^\dagger\\
\tilde{\alpha} f_{-\sigma}\\
b_\alpha^\dagger\\
\tilde{\alpha} b_{-\sigma}
\end{array} \right)
\end{eqnarray}
is a four-component spinor, and as defined in the main text $\gamma_0=diag[1,1,1,-1]$, and $\tilde{\alpha}=sgn(\alpha)$.

Under a super-rotation of the spinors $\Psi_\alpha\rightarrow g \Psi_\alpha$, the spin operator transforms as:
\begin{eqnarray}
S_{\alpha\beta}= \Psi^\dagger_{\alpha} \gamma_0 \Psi_{\beta}\rightarrow  \Psi^\dagger_{\alpha} g^\dagger\gamma_0 g\Psi_{\beta},
\end{eqnarray}
so for the spin to be invariant the transformation $g$ should satisfy:
\begin{eqnarray}
g^\dagger \gamma_0 g=\gamma_0,
\end{eqnarray}
which is essentially the unitarity condition to the transformation after taking appropriate care of the commutativity of the bosons.

The most general transformation $g$ can be obtained by exponentiation of the generators of the algebra introduced in Eq. \ref{DefOp}. Exponentiation of the even (or commuting) part of the algebra  gives:
\begin{eqnarray}
g_E=\left( \begin{array}{cccc}
u&v&0&0\\
-\bar{v}&\bar{u}&0&0\\
0&0&x&0\\
0&0&0&\bar{x}
\end{array} \right),
\end{eqnarray}
where the parameters $u$, $v$ and $x$ are complex numbers satisfying $|u|^2+|v|^2=1$ and $|x|^2=1$. Note the $SU(2)$ and $U(1)$ substructure of this transformation for the fermionic and bosonic parts of the spinor $\Psi_\sigma$, respectively.

Exponentiating the odd (or anticommuting) part of the algebra we find:
\begin{eqnarray}
g_A=\left( \begin{array}{cccc}
1+\frac{\alpha\bar{\alpha}}{2}&0&-\bar{\alpha}&0\\
0&1+\frac{\alpha\bar{\alpha}}{2}&0&-\alpha\\
\alpha&0&1-\frac{\alpha\bar{\alpha}}{2}&0\\
0&-\bar{\alpha}&0&1-\frac{\alpha\bar{\alpha}}{2}
\end{array} \right),
\end{eqnarray}
and
\begin{eqnarray}
g_B=\left( \begin{array}{cccc}
1-\frac{\beta\bar{\beta}}{2}&0&0&-\bar{\beta}\\
0&1-\frac{\beta\bar{\beta}}{2}&\beta&0\\
0&-\bar{\beta}&1+\frac{\beta\bar{\beta}}{2}&0\\
-\beta&0&0&1+\frac{\beta\bar{\beta}}{2}
\end{array} \right),
\end{eqnarray}
where the parameters $\alpha$ and $\beta$ are complex Grassmann numbers.

The most general transformation can be obtained by the composition of the three transformations above:
\begin{eqnarray}
g=g_E g_A g_B,
\end{eqnarray}
which satisfies the condition $g^\dagger \gamma_0 g=\gamma_0$, as required.

\subsection{Derivation of the Casimir}\label{AppSSNwa} We now follow
Nwachuku\cite{Nwa} in order to derive the second Casimir of $SP (N)$, 
(where $N$ is an even number), given in
Eq.~\ref{NwaEq}. Each irreducible representation of $SP (N)$
is characterized by the set of integers
$(f_{N/2},f_{N/2-1},...,f_1)$. This set of numbers
corresponds to the number of boxes in each row of the respective
Young tableau, starting from the topmost and longest row, which has $f_{N/2}$
boxes. These numbers also 
correspond to the eigenvalues of the $N/2$
Cartan (diagonal) generators in the highest state of the
representation. 
Following \cite{Nwa}, if we define:
\begin{eqnarray}
&&\lambda_i=f_i+N/2+i, \hspace{0.5cm}\text{for $i\geq0$}\\
&&\lambda_{-i}=-\lambda_i+N, \hspace{0.5cm}\text{for $i>0$}\\
&&\rho_i=N/2+i,
\end{eqnarray}
where we take $f_{0}=0$, the second Casimir is written:
\begin{eqnarray}
C_2=\sum_{i=-N/2}^{N/2}(\lambda_i^2-\rho_i^2).
\end{eqnarray}

For an L-shaped tableau with width $w$ and height $h$, we have:
\begin{equation}
  f_i=\begin{cases}
    w, & \text{$i=\frac{N}{2}$},\\
    1, & \text{$\frac{N}{2}-1\geq i \geq \frac{N}{2}-h+1$},\\
    0, & \text{$i\leq \frac{N}{2}-h$},
  \end{cases}
\end{equation}
%\vspace{0.4cm}

\newpage

therefore:
\begin{equation}
  \lambda_i=\begin{cases}
    w+N, & \text{$i=\frac{N}{2}$},\\
   -w, & \text{$i=-\frac{N}{2}$},\\
    1+N/2+i, & \text{$\frac{N}{2}-1\geq i \geq \frac{N}{2}-h+1$},\\
     -1+N/2+i, & \text{$-|\frac{N}{2}-1|\leq i \leq -|\frac{N}{2}-h+1|$},\\
    0, & \text{otherwise},
  \end{cases}
\end{equation}
and
\begin{equation}
  \rho_i=\begin{cases}
    N, & \text{$i=\frac{N}{2}$},\\
   0, & \text{$i=-\frac{N}{2}$},\\
    N/2+i, & \text{$\frac{N}{2}-1\geq i \geq \frac{N}{2}-h+1$},\\
    N/2+i, & \text{$-|\frac{N}{2}-1|\leq i \leq -|\frac{N}{2}-h+1|$},\\
    0, & \text{otherwise},
  \end{cases}
\end{equation}
so we can perform the sum:
\begin{eqnarray}
C_2&=& w^2+4Nw + w^2\\&+&\sum_{i=N/2-h+1}^{N/2-1}(1+N+2i)\nonumber\\&+&\sum_{i=-N/2+1}^{-N/2+h-1}(1-N-2i),\nonumber\\
&=&2(w+h)(N+w-h) + 4(h-N/2) -2,\nonumber
\end{eqnarray}
where the sums were evaluated as sums of arithmetic progressions. This is Eq.~\ref{NwaEq} in the main text. Identifying $Q=w+h-1$ and $Y=h-w$, we have
\begin{eqnarray}\label{l}
C_2&=& 2 (Q+1) (N-Y)+2 (Y+Q+1-N)-2\cr
&=& 2 (Q+1) (N-Y)-2N + 2 (Y+Q)\cr
&=& 2Q(N+1-Y),
\end{eqnarray}
which is Eq.~\ref{NwaEq2}, similar to the form discussed in Coleman et al.\cite{Col1} in the case of an SU(N) generalization of the supersymmetric spin.

\begin{widetext}
\subsection{Operator form of the Casimir}
\label{AppA4}
Now we relate the magnitude of the spin $\mathbf{S}^2$ with the Casimir computed above. We start computing:
\begin{eqnarray}
\mathbf{S}^2&=&\sum_{\alpha\beta} S_{\alpha\beta}S_{\beta\alpha},
\end{eqnarray}
with $S_{\alpha\beta}$ is defined in Eq.~\ref{S}. Taking the operator products and relabeling the summed indexes we find:
\begin{eqnarray}
\mathbf{S}^2&=& 2( 
f_\alpha^\dagger f_\beta f_\beta^\dagger f_\alpha 
+ b_\alpha^\dagger b_\beta b_\beta^\dagger b_\alpha
+2 f_\alpha^\dagger f_\beta b_\beta^\dagger b_\alpha\cr
&-&
\tilde{\alpha}\tilde{\beta} f_\alpha^\dagger f_\beta f_{-\alpha}^\dagger f_{-\beta} 
-\tilde{\alpha}\tilde{\beta}b_\alpha^\dagger b_\beta b_{-\alpha}^\dagger b_{-\beta}
 -2\tilde{\alpha}\tilde{\beta} f_\alpha^\dagger f_\beta b_{-\alpha}^\dagger b_{-\beta}),
\end{eqnarray}
where we have used summation convention for the repeated spin
indices. 
The first line of this expression is the Casimir for the $SU(N)$ spin,
while the second line introduces the additional cross-terms that result
from the symplectic form of the spin. 

Expanding the first line, we have
\begin{eqnarray}\label{l}
f\dg_{\alpha }f_{\beta }f\dg_{\beta }f_{\alpha }&=& n_{F} (N-n_{F})+n_{F}\cr
b\dg_{\alpha }b_{\beta }b\dg_{\beta }b_{\alpha }&=& n_{B}
(N+n_{B})-n_{B}\cr
f\dg_{\alpha }f_{\beta }b\dg_{\beta }b_{\alpha }&=& \theta \dg \theta -n_{F}
\end{eqnarray}
so that 
\begin{eqnarray}\label{termA}
f_\alpha^\dagger f_\beta f_\beta^\dagger f_\alpha 
+ b_\alpha^\dagger b_\beta b_\beta^\dagger b_\alpha
+2 f_\alpha^\dagger f_\beta b_\beta^\dagger b_\alpha = (n_{B}+n_{F})
(N+n_{B}-n_{F})+ 2 \theta \dg  \theta - (n_{B}+n_{F})
\end{eqnarray}
Now expanding the additional cross terms, 
\begin{eqnarray}\label{l}
-\tilde{\alpha}\tilde{\beta} f_\alpha^\dagger f_\beta
f_{-\alpha}^\dagger f_{-\beta} &=&  n_{F} - \tilde{\alpha}\tilde{\beta
}f\dg_{\alpha }f\dg_{-\alpha } f_{-\beta }f_{\beta }
=  n_{F}- 4 \psi \dg \psi ,\cr
-\tilde{\alpha}\tilde{\beta}b_\alpha^\dagger b_\beta b_{-\alpha}^\dagger b_{-\beta}
&=&  n_{B} - \tilde{\alpha}\tilde{\beta}b_{\alpha }\dg  b_{-\alpha}\dg  
b_{-\beta }b_{\beta }
 =  n_{B},\cr
 -2\tilde{\alpha}\tilde{\beta} f_\alpha^\dagger f_\beta
 b_{-\alpha}^\dagger b_{-\beta}&=& - 2\tilde{\alpha }\tilde{\beta }
 f\dg_{\alpha } b\dg _{-\alpha }b_{-\beta }f_{\beta } = - 2 \eta \dg \eta
\end{eqnarray}
where we have used the fact that  the bosonic pairs vanish 
$\sum_{\beta }\tilde{\beta }b_{-\beta }b_{\beta }=0$. Combining the
additional cross terms, we have 
\begin{eqnarray}\label{termB}
-\tilde{\alpha}\tilde{\beta} f_\alpha^\dagger f_\beta f_{-\alpha}^\dagger f_{-\beta} 
-\tilde{\alpha}\tilde{\beta}b_\alpha^\dagger b_\beta b_{-\alpha}^\dagger b_{-\beta}
 -2\tilde{\alpha}\tilde{\beta} f_\alpha^\dagger f_\beta
 b_{-\alpha}^\dagger b_{-\beta}) = n_{F}+n_{B}- 4 \psi\dg\psi -2\eta
 \dg  \eta.
\end{eqnarray}
Combining (\ref{termA}) and (\ref{termB}), noting that the remainder
terms $\pm (n_{B}+n_{F})$ cancel one-another, we obtain
\begin{eqnarray}\label{spincasimiry}
\mathbf{S}^2&=&2\left[
(\hat{n}_B+\hat{n}_F)(N+\hat{n}_B-\hat{n}_F)
-4\hat{\psi}^\dagger\hat{ \psi}+2
 \hat{\theta}\dg  \hat{\theta} -2\hat{\eta}^\dagger \hat{\eta}
\right],
\end{eqnarray}
so we can identify:
\begin{eqnarray}
\mathbf{S}^2&=&2 \hat{Q}(N+1-\hat{Y}),
\end{eqnarray}
where
\begin{eqnarray}
\hat{Q}&=&\hat{n}_F+\hat{n}_B,\\ 
\hat{Y}&=&\hat{n}_F-\hat{n}_B+1 +\frac{4\hat{\psi}^\dagger \hat{\psi}-2 \hat{\theta}\dg   \hat{\theta} +2\hat{\eta}^\dagger \hat{\eta}}{\hat{Q}}.
\end{eqnarray}

\subsection{Sum rule for spin and super-Casimir}\label{}
\label{AppA5}
The second Casimir invariant $\chi^{2}$ of the $SU (2|1)$ group can be written \cite{hopkinson}
\begin{equation}\label{}
\chi ^{2} 
%= {\rm Str}[(gX)^{2}]
= {\rm Tr}[X m X],
\end{equation}
where $X\equiv X_{ab}$ is the three-dimensional matrix formed out of
the Hubbard operators and $m={\rm diag} (1,1,-1)$. If we expand this
result we obtain
\begin{equation}\label{}
\chi ^{2}= X_{\alpha\beta}X_{\beta\alpha}
-
[X_{\alpha 0},X_{0\alpha}]- (X_{00})^{2},
\end{equation}
with an implied summation over the repeated indices $\alpha, \beta =
\pm$.  Substituting for the Hubbard operators using (\ref{thehubbard})
we obtain
\begin{eqnarray}\label{supercas}
\chi ^{2} &=& X_{++}^{2} + X_{--}^{2} -X_{00}^{2}+
 \{X_{+-},X_{-+}\} -[X_{+0},X_{0+}]-[X_{-0},X_{0-}]\cr
&=& [(n_{F}-n_{B})/2]^{2} + [(N-n_{F}-n_{B})/2]^{2} - \hat
n_{B}^{2}
+ \{\hat \psi
\dg,\hat \psi   \}-[\hat \theta \dg ,\hat  \theta ]/2
-[\hat \eta  ,\hat  \eta\dg ]/2
\cr
&=& 
N^{2}/4 - (n_{F}+n_{B}) (N -n_{F}+n_{B})/2 
+ \{\psi \dg , \psi\}- [\theta \dg ,\theta ]/2 + [\eta \dg ,\eta ]/2.
\end{eqnarray}
Using the Hubbard algebra of the $SU (2|1)$ generators (\ref{hubbardalgebra}), we have 
\begin{eqnarray}\label{l}
\{\psi \dg ,\psi \}&=&
 2 \psi \dg  \psi + [\psi ,\psi \dg ] 
= 2 \psi \dg  \psi 
+ [X_{-+},X_{+-}] 
=  2 \psi \dg  \psi +X_{--}-X_{++},\cr
[\theta \dg ,\theta ]/2 &=& \theta \dg \theta - \{\theta ,\theta \dg \}/2
= \theta \dg \theta - \{X_{0+},X_{+0}\}=
\theta \dg  \theta - (X_{00}+X_{++}),\cr
[\eta \dg ,\eta ]/2 &=& \eta \dg \eta - \{\eta ,\eta \dg \}
= \eta \dg \eta - \{X_{-0},X_{0-}\}=
\eta \dg  \eta- (X_{--}+X_{00}),
\end{eqnarray}
Adding up these expressions, we find that
\begin{equation}\label{}
\{\psi \dg , \psi\}- [\theta \dg ,\theta ]/2 + [\eta \dg ,\eta ]/2 = 
2\psi \dg  \psi- \theta \dg \theta + \eta \dg \eta .
\end{equation}
Inserting this into (\ref{supercas}) we obtain
\begin{equation}\label{}
\chi^{2} = 
N^{2}/4 - (n_{F}+n_{B}) (N -n_{F}+n_{B})/2
+2\psi \dg  \psi- \theta \dg \theta + \eta \dg \eta .
\end{equation}
Finally, comparing with (\ref{spincasimiry}), we obtain
\begin{equation}\label{}
\chi^{2} = N^{2}/4 - {\bf S}^{2}/4.
\end{equation}
The corresponding sum rule 
\begin{equation}\label{}
N^{2}/4 = \chi^{2}+ {\bf S}^{2}/4.
\end{equation}
expresses the fact that 
sum of the pair/charge fluctuations and the spin fluctuations is a constant. 
\end{widetext}

%%%%%%%%%%%%%%%%%%%%%%%%%%%%%%%%%%%%%%%%%%%%%%

\section{Computation of the Fermionic part of the free energy}\label{AppFE}

The fermionic part of the solution reduces to two decoupled impurity problems. Here we explicit the calculation for a single impurity. The partition function for a single impurity can be written as:
\begin{eqnarray}
Z_F&=&\int \mathcal{D}\mu_F e^{-S_F}, \hspace{0.5cm} \mathcal{D}\mu_F = \mathcal{D}[c, f, v,\lambda_{F}]\nonumber,
\end{eqnarray}
already transforming from imaginary time to Matsubara frequencies, the action can be written as:
\begin{eqnarray}
S_F &&=\sum_n \left[\sum_{\bk\sigma} c^\dagger_{\bk\sigma}(-i\omega_n +\delta_{\bk})c_{\bk\sigma}\right.\nonumber\\ && \left.+\sum_{\sigma} f^\dagger_{\sigma}(-i\omega_n+\lambda_{F}) f_{\sigma}\right.\nonumber\\&&\left. +\sum_{\sigma}\left(\sum_{\bk} f_{\sigma}^\dagger v c_{\bk\sigma}+ h.c.\right)\right]\nonumber\\&&+\beta N \sum_{\bk}\frac{ |v|^2}{J_K}-\beta N \lambda_{F} q_F.
\end{eqnarray}

We start by integrating out the conduction electrons, taking into account their effect in the self energy of the fermions that compose the spin. The effective fermion propagator can be written as:
\begin{eqnarray}
G_f =[(G_f^0)^{-1} -\Sigma_f]^{-1}, 
\end{eqnarray}
where $(G_f^{0})^{-1}=i\omega_n-\lambda_F$ is the bare f-fermion propagator, and 
\begin{eqnarray}
\Sigma_f = \sum_\bk |v|^2  G_{c\bk}^0,
\end{eqnarray}
is the f-fermion free energy, where $(G_{c\bk}^0)^{-1}= i\omega_n-\delta_\bk$ is the bare conduction electron propagator. We evaluate the sum over $\bk$ in $\Sigma_f$ as an integral over energy with a constant density of states. Analytically continuing the Matsubara frequencies to the real axis ($\omega_n\rightarrow \omega \pm i \delta$):
\begin{eqnarray}
\Sigma_f &=& |v|^2\sum_\bk \frac{1}{\omega\pm i\delta -\epsilon_\bk}\nonumber\\
&=&|v|^2\int_{-D}^D \rho(\epsilon) d\epsilon \left(\frac{1}{w-\epsilon}\mp i \pi \delta(\omega-\epsilon)\right),\nonumber\\
&=& -i\Gamma\Theta(D-|\omega|) sgn(\tilde{\omega}),
\end{eqnarray}
where $\Gamma=\pi \rho_0 |v|^2$, $D$ the bandwidth, $\rho_0$ the constant density of states, $\Theta(x)$ the Heaviside step function. Here $\tilde{\omega}$ indicates the imaginary part of the frequency.

Now the fermionic part of the free energy reads:
\begin{eqnarray}
&&S_{F}=\sum_{n \sigma} f^\dagger_{\sigma}(i\omega_n)(-i\omega_n+\lambda_{F}+i\Gamma_n) f_{\sigma}(i\omega_n)\nonumber\\&&\hspace{0.8cm}+\beta\frac{N |v|^2}{J_K}-\beta N \lambda_{F} q_F,
\end{eqnarray}
where $\Gamma_n=\Gamma\Theta(D-|\omega_n|) sgn(\tilde{\omega}_n)$. Now we can integrate out the f-fermions and write an effective action at the saddle point values of $v$ and $\lambda_F$ (to be determined by extremization of the free energy, see main text):
\begin{eqnarray}
S_F^{Eff}= -\sum_{n\sigma}\log[-i\omega_n+\lambda_F+i\Gamma_n]\nonumber\\+\beta\frac{N |v|^2}{J_K}-\beta N \lambda_{F} q_F.
\end{eqnarray}

The sum over Matsubara frequencies can be performed as an integral in the complex plane weighted by the Fermi distribution function $f(z)=(e^{\beta z}-1)^{-1}$. Note that the integral involves a branch-cut:
\begin{eqnarray}
&&\sum_{n\sigma}\log[-i\omega_n+\lambda_F+i\Gamma\Theta(D-|\omega_n|) sgn(\tilde{\omega}_n)]\nonumber\\&=&
\frac{\beta N}{2\pi i}\int_C dz \log[-z+\lambda_F+i\Gamma \Theta(D-|z|)sgn(\tilde{z})] f(z)\nonumber\\
&=&\frac{\beta N}{2\pi i} \left[\int_{-D}^D dz f(z)\log[-z+\lambda_F -i\Gamma]\right.\nonumber\\ &&\left.+\int_{D}^{-D}dzf(z)\log[-z+\lambda_F+i\Gamma]\right],
\end{eqnarray}
which simplifies to:
\begin{eqnarray}
-\frac{\beta N}{\pi} \int_{-D}^D dz f(z) Im[\log[-z+\lambda_F+i\Gamma]].
\end{eqnarray}

In the zero temperature limit the Fermi function sets the upper limit of the integral to zero. Evaluating the integral we find the free energy:
\begin{eqnarray}
\frac{F_F}{N}&=&\frac{1}{\pi} Im\left[ (\lambda_F+i\Gamma) \ln \left(\frac{\lambda_F+i\Gamma}{D e}\right)\right]\nonumber\\&&+\frac{|v|^2}{J_K}-\lambda_F q_F,
\end{eqnarray}
that can be rewritten as:
\begin{eqnarray}
\frac{F_F}{N}&=&\frac{1}{\pi} Im\left[ \xi_F \ln \left(\frac{\xi_F}{eT_K e^{i\pi q_F}}\right)\right],
\end{eqnarray}
once we define
\begin{eqnarray}
\xi_F&=&\lambda_F+i\Gamma,
\end{eqnarray}
and the Kondo temperature
\begin{eqnarray}
T_K=D e^{-1/\rho_0 J_K}.
\end{eqnarray}

%%%%%%%%%%%%%%%%%%%%%%%%%%%%%%%%%%%%%%%%%%%%%%

\section{Computation of the bosonic part of the free energy}\label{AppBE}

From the main text we have that the bosonic part of the partition function is:
\begin{eqnarray}
Z_B&=&\int \mathcal{D}\mu_B e^{-S_B}, \hspace{0.5cm} \mathcal{D}\mu_B = \mathcal{D}[b, g, \lambda_{B}]\nonumber,
\end{eqnarray}
already transforming from imaginary time to Matsubara frequencies, the action can be written as:
\begin{eqnarray}
S_B&=&\sum_{n\sigma} \Psi_{B\sigma}^\dagger(i\nu_n) L_B(i\nu_n) \Psi_{B\sigma}(i\nu_n) \nonumber\\&&+\beta N \frac{ |g|^2}{J_H}-2\beta N\lambda_{B}( q_B+1/2),
\end{eqnarray}
where
\begin{eqnarray}
L_B(i\nu_n)=\left( \begin{array}{cc}
-i\nu_n+\lambda_B& g   \\
\bar{g} & i\nu_n +\lambda_B \\
\end{array} \right),
\end{eqnarray}
\begin{eqnarray}
\Phi_{B\sigma}(i\nu_n)=\left( \begin{array}{c}
b_{1\alpha} (i\nu_n)\\
\tilde{\sigma} b_{2-\sigma}^\dagger(-i\nu_n)\\
\end{array} \right).
\end{eqnarray}

Integrating out the bosons and taking the saddle point value of $\lambda_B$ and $g$, which will be determined by the extremization of the free energy with respect to these parameters, we can write:
\begin{eqnarray}
Z_B&=& e^{-S_B^{Eff}},
\end{eqnarray}
where
\begin{eqnarray}
S_B^{Eff} &=& \sum_{n\sigma} \log [Det[L_B(i\nu_n)]]+\beta N \frac{ |g|^2}{J_H}\nonumber\\&&-2\beta N\lambda_{B}( q_B+1/2),\\
&=& N \sum_{n,x=\pm} \log[E_B^x-i\nu_n]+\beta N \frac{ |g|^2}{J_H}\nonumber\\&&-2\beta N\lambda_{B}( q_B+1/2),\nonumber
\end{eqnarray}
where
\begin{eqnarray}
E_B^\pm = \pm \sqrt{\lambda_B^2-|g|^2}.
\end{eqnarray}

The sum over Matsubara frequencies can be written in terms of an integral over the imaginary plane weighted by the bosonic distribution function $n(z)=(e^{\beta z}-1)^{-1}$:
\begin{eqnarray}
\sum_{n,x=\pm} \log[E_B^x-i\nu_n]= -\beta N\sum_{x=\pm} \int_C\frac{dz}{2\pi i} \log[E_B^x-z] n(z).\nonumber\\
 \end{eqnarray}
 
In the zero temperature limit:
\begin{eqnarray}
\sum_{n\sigma} \log [Det[L_B(i\nu_n)]] \xrightarrow{T\rightarrow 0} N \sum_{x=\pm} (-E_B^x)\Theta(-E_B^x),\nonumber\\
\end{eqnarray}
where $\Theta(x)$ is the Heaviside step function, so that the bosonic part of the free energy reads:
\begin{eqnarray}
\frac{F_B}{N}&=&\sqrt{\lambda_B^2-|g|^2} + \frac{ g^2}{J_H}-2\lambda_{B} (q_B+1/2).
\end{eqnarray}

%%%%%%%%%%%%%%%%%%%%%%%%%%%%%%%%%%%%%%%%%%%%%%

\section{Computation of the fluctuations of the fermionic hybridization}\label{AppFlu}

In this appendix we define and compute $\chi_{cb}(i\omega_n)$. From the main text we have:
\begin{eqnarray}
\chi_{cb}(i\omega_r)= -\frac{1}{\beta}\sum_{\bk \bk' m}G_{b} (i\omega_{r}+i\omega_m) G_{\bk\bk'}(i\omega_m)
\end{eqnarray}
where,
\begin{eqnarray}
G_{b}(i\nu_n)&=&(i\nu_n-\xi_B)^{-1}
\end{eqnarray}
is the bosonic propagator, with $\xi_B=\sqrt{\lambda_B^2-|g|^2}$, and
\begin{eqnarray}
G_{\bk\bk'}(i\omega_n)&=&G_\bk^0(i\omega_n)\delta_{\bk\bk'}\nonumber\\&+&|v|^2G_\bk^0(i\omega_n)G_f(i\omega_n)G_{\bk'}^0(i\omega_n),
\end{eqnarray}
with the propagators defined in Appendix~\ref{AppFE}.

Evaluating the sum over momenta:
\begin{eqnarray}
\sum_{\bk\bk'}G_{\bk\bk'}(i\omega_n)&=&\sum_\bk G_\bk^0(i\omega_n)\nonumber\\&+&|v|^2G_f(i\omega_n)\left(\sum_\bk G_{\bk}^0(i\omega_n)\right)^2,
\end{eqnarray}
where
\begin{eqnarray}
\sum_\bk G_\bk^0(i\omega_n)=-i \pi \rho_0 sgn(\omega_n),
\end{eqnarray}
as computed in the evaluation of the fermionic part of the free energy. In the infinite bandwidth limit the sum over $\bk$ can be written as:
\begin{eqnarray}
\sum_{\bk\bk'} G_{\bk\bk'}(i\omega_n)&=&- i \pi \rho_0 sgn(\omega_n)\nonumber\\&-&\frac{\pi \rho_0 \Gamma}{i\omega_n- \lambda_F+i\Gamma sgn (\omega_n)}
\end{eqnarray}
where $\rho_0$ is a constant density of states and $\Gamma=\pi\rho_0|v|^2$ as before.

Back to the computation of $\chi_{cb}$:
\begin{eqnarray}
\chi_{cb}(i\omega_r)&=&\chi^1_{cb}(i\omega_r)+\chi^2_{cb}(i\omega_r),
\end{eqnarray}
where
\begin{eqnarray}
\chi_{cb}^1(i\omega_r)&=& \frac{1}{\beta}\sum_{ m}\frac{i \pi \rho_0 sgn(\omega_m)}{i\omega_{m}+i\omega_r-\xi_B}\nonumber\\
&=&\frac{i \pi \rho_0}{2 \pi i}\oint dz f(z) \frac{sgn(\tilde{z})}{z+i\omega_{r}-\xi_B}\nonumber\\
&=&\frac{i \pi \rho_0}{2 \pi i}\left[ 2\pi i f(\lambda_B-i\omega_r)(-1)\right.\nonumber\\&+&\int_{-D}^D dz f(z) \frac{(-1)}{z+i\omega_r-\xi_B}\nonumber\\&+&\left.\int_{D}^{-D}dz f(z)\frac{(+1)}{z+i\omega_r-\xi_B}\right],\nonumber\\
\end{eqnarray}
where $\tilde{z}=Im(z)$ and $f(\lambda_B-i\omega_r)=-n(\lambda_B)$. In the zero temperature limit $f(z)\rightarrow \theta(-z)$ and $n(\lambda_B>0)\rightarrow0$, so:
\begin{eqnarray}
\chi_{cb}^1(i\omega_r)&=& - \rho_0 \log\left(\frac{-\xi_B+i\omega_r}{-\xi_B+i\omega_r-D}\right).
\end{eqnarray}

The second part of $\chi_{cb}(i\omega_r)$:
\begin{widetext}
\begin{eqnarray}
\chi_{cb}^2(i\omega_r)&=& \frac{1}{\beta}\sum_{ m}\frac{\pi \rho_0 \Gamma}{i\omega_m- \lambda_F+i\Gamma sgn (\omega_m)}\frac{1}{i\omega_{m}+i\omega_r-\xi_B}\nonumber\\
\end{eqnarray}
can be computed in analogous fashion:
\begin{eqnarray}
\chi_{cb}^2(i\omega_r)&=&\frac{ \rho_0 \Gamma}{2 i}\frac{1}{i\omega_r-\xi_B+\lambda_F+i\Gamma}\left[ Log\left(\frac{-\lambda_F-i\Gamma}{-\lambda_F-i\Gamma-D}\right)-Log\left(\frac{-\xi_B+i\omega_r}{-\xi_B+i\omega_r-D}\right)\right] \nonumber\\&& -\frac{ \rho_0 \Gamma}{2 i} \frac{1}{i\omega_r-\xi_B+\lambda_F-i\Gamma}\left[ Log\left(\frac{-\lambda_F+i\Gamma}{-\lambda_F+i\Gamma-D}\right)-Log\left(\frac{-\xi_B+i\omega_r}{-\xi_B+i\omega_r-D}\right)\right].\nonumber\\
\end{eqnarray}

Continuing to real frequencies $\omega_r\rightarrow \omega-i\delta$, writing $1/J_K=-\rho_0 Log|(\lambda+i\Gamma)/D|$, in the infinite bandwidth limit:
\begin{eqnarray}
\chi_{cb}(\omega -i\delta)-\frac{1}{J_K}&=&+\rho_0 Log\left| \frac{\lambda_F+i\Gamma}{\xi_B-\omega+i\delta}\right|-i\pi\rho_0\Theta(\omega-\lambda)+\frac{ \rho_0 \Gamma}{2 i}\frac{1}{\omega-i\delta-\xi_B+\lambda_F+i\Gamma} Log\left(\frac{\lambda_F+i\Gamma}{\xi_B-\omega+i\delta}\right) \nonumber\\&& -\frac{ \rho_0 \Gamma}{2 i} \frac{1}{\omega-i\delta-\xi_B+\lambda_F-i\Gamma} Log\left(\frac{\lambda_F-i\Gamma}{\xi_B-\omega+i\delta}\right).
\end{eqnarray}

In the transition line, where $\xi_B=\xi_F\Rightarrow \sqrt{\lambda_B^2-g^2}=\lambda_B=\lambda_F=\lambda$, we have the simplified form:
\begin{eqnarray}
\chi_{cb}(\omega -i\delta)-\frac{1}{J_K}&=&+\rho_0 Log\left| \frac{\lambda+i\Gamma}{\lambda-\omega+i\delta}\right|-i\pi\rho_0\Theta(\omega-\lambda)\nonumber\\&&+
\frac{ \rho_0 \Gamma}{2 i}\frac{1}{\omega^2+\Gamma^2}\left[ (\omega-i\Gamma)Log\left(\frac{\lambda+i\Gamma}{\lambda-\omega+i\delta}\right)-(\omega+i\Gamma)Log\left(\frac{\lambda-i\Gamma}{\lambda-\omega+i\delta}\right)\right],\nonumber\\
\end{eqnarray}
rewriting,
\begin{eqnarray}
\chi_{cb}(\omega -i\delta)&-&\frac{1}{J_K}= \rho_0 \omega Re\left[ \frac{ Log(\lambda+i\Gamma)}{\omega+i\Gamma}\right]-\frac{\rho_0  \omega ^2}{\Gamma ^2+\omega ^2}\log (\lambda -\omega + i\delta),
\end{eqnarray}
which is the form of $\chi_{cb}$ discussed in the main text.

\end{widetext}

%%%%%%%%%%%%%%%%%%%%%%%%%%%%%%%%%%%%%%%%%%%%%%


\begin{thebibliography}{99}
\bibitem{FPtrick} One way to avoid the constraint is to use the Fedotov-Popov trick \cite{Fed}, by the introduction of an imaginary chemical potential, but note that this approach makes the Hamiltonian non-Hermitian and the convexity arguments used in variational approaches to determine the solution as an upper bound to the real ground state energy cannot be applied in this case.
\end{thebibliography}
\end{document}